\documentclass{article}
\usepackage[left=0.9in,right=0.9in,top=1.2in,bottom=0.9in]{geometry}

\usepackage{booktabs} 
\usepackage{enumitem}
\usepackage{hyperref} 
\usepackage{balance} 
\usepackage{pifont} 
\usepackage{tabu}
\usepackage[table]{xcolor}
\usepackage{url} 
\usepackage{array} 
\usepackage{tikz}

\usetikzlibrary{shapes,positioning}

\newcommand\ytl[3]{
\parbox[b]{8em}{\hfill{\color{black}#1}~$\cdots$~}
\makebox[0pt][c]{$\bullet$}\vrule\quad \parbox[c]{5cm}{\vspace{5pt}\color{black!80}\raggedright #2.\\[5pt]}
\makebox[0pt][c]{}\vrule\quad 
\parbox[c]{6cm}{\vspace{5pt}\color{black!80}\raggedright #3.\\[5pt]}\\[-3pt]}

\definecolor{LightCyan}{rgb}{0.88,1,1}
\definecolor{NiceGold}{RGB}{218,165,32}
\definecolor{checked}{RGB}{0,128,0}

\usepackage{bbding}
\usepackage[numbers,sort]{natbib}

\usepackage{authblk}

\title{The Query Translation Landscape: a Survey}

\author[1]{Mohamed Nadjib Mami}
\author[2,1]{Damien Graux}
\author[3]{Harsh Thakkar}
\author[1]{Simon Scerri}
\author[4,5]{Sören Auer}
\author[1,3]{Jens Lehmann}

\affil[1]{Enterprise Information Systems, Fraunhofer IAIS, St.~Augustin \& Dresden, Germany}
\affil[2]{ADAPT Centre, Trinity College of Dublin, Ireland}
\affil[3]{Smart Data Analytics group, University of Bonn, Germany}
\affil[4]{TIB Leibniz Information Centre for Science and Technology, Germany}
\affil[5]{L3S Research Center, Leibniz University of Hannover, Germany}

\date{
    October 2019
}

\begin{document}

\maketitle

\begin{abstract}
Whereas the availability of data has seen a manyfold increase in past years, its value can be only shown if the data variety is effectively tackled ---one of the prominent Big Data challenges. The lack of data interoperability limits the potential of its collective use for novel applications. Achieving interoperability through the full transformation and integration of diverse data structures remains an ideal that is hard, if not impossible, to achieve. Instead, methods that can simultaneously interpret different types of data available in different data structures and formats have been explored.
On the other hand, many query languages have been designed to enable users to interact with the data, from relational, to object-oriented, to hierarchical, to the multitude emerging NoSQL languages. Therefore, the interoperability issue could be solved not by enforcing physical data transformation, but by looking at techniques that are able to query heterogeneous sources using one uniform language. Both industry and research communities have been keen to develop such techniques, which require the translation of a chosen 'universal' query language to the various data model specific query languages that make the underlying data accessible.
 
In this article, we survey more than forty query translation methods and tools for popular query languages, and classify them according to eight criteria.
In particular, we study which query language is a most suitable candidate for that 'universal' query language. Further, the results enable us to discover the weakly addressed and unexplored translation paths, to discover gaps and to learn lessons that can benefit future research in the area.
\end{abstract}

\section*{Introduction}
Query languages have come a long way during the last few decades. The first database query language, SQL, was formally introduced in the early seventies~\cite{chamberlin1974sequel} following the earlier proposed and well-received relational model~\cite{codd1970relational}. SQL has influenced the design of dozens query languages, from several SQL dialects, to object-oriented, graph, columnar, and the various NoSQL languages. These query languages are implemented and used in an unprecedented variety of storage and data management systems. 
In order to leverage the advantages of these solutions, companies and institutions are choosing to store their data in different representations, a phenomenon known as \textit{Polyglot Persistence}~\cite{sadalage2013nosql}. As a result, large data repositories with heterogeneous data sources are being generated (also known as \emph{Data Lakes}~\cite{2010dixon}), exposing various query interfaces to the user.
Integrating this heterogeneous data (Big Data Variety~\cite{laney2012deja}) into a unified format and system, as has historically been the case with e.g., data warehouses, is nowadays becoming irrelevant. This is because (1) data is very large in size (Big Data Volume), (2) companies are less likely to sacrifice data freshness especially with the advances in streaming and IoT technologies (Big Data Velocity).

On the other hand, while computer scientists were looking for the holy grail of data representation and querying in the last decades, it is meanwhile accepted that no optimal data storage and query paradigm exist. Instead, different storage and query paradigms have different characteristics especially in terms of representation and query expressivity and scalability. Different approaches balance differently between expresivity and scalability in this regard. While SQL, for example, comprises a sophisticated data structuring and very expressive query language, NoSQL trades schema and query expressivity for scalability. As a result, since no optimal representation exists, different storage and query paradigms have their right to exist based on the requirements of various usecases. 

With the resulted high variety, the challenge is how can the collected data sources be accessed in a uniform ad hoc way. Learning the syntax of their respective query languages is counterproductive as these query languages may substantially differ in both their syntax and semantics. A plausible approach is to develop means to map and translate between different storage and query paradigms. One way to achieve this is by leveraging the existing query translators, and building wrappers that allow the conversion of a query in a unique language to the various query languages of the underlying data sources.
This has stressed the need for a better understanding of the translation methods between the query languages.

The topic covered in this survey, namely Query Translation, is horizontal to and directly concerns many Computer Science domains, from Information Retrieval, Databases, Data Integration, Data Analytics, Polyglot Persistence to Data Publishing and Archiving. Thus, the topic can be of interest to a broad audience; from as specific as researchers in Query Translation topics, to as general as users who solely interact with an existing system using those query languages and needing to transition from one language to another.

\paragraph{Related Surveys.} Several studies investigating query translation methods exist in the literature. They typically tackle pair-wise translation methods between two specific types of query languages, e.g., \cite{krishnamurthy2003xml} surveys XML languages-to-SQL query translations, \cite{michel2014survey,spanos2012bringing,sahoo2009survey} surveys SPARQL-to-SQL query translations.
However, to the best of our knowledge, no survey has tackled the problem of universal translation across several query languages. 

\paragraph{Contributions.} In this survey article we take a broader view over the query translation landscape. We consider existing query translation methods that target many widely-used and standardized query languages. Those include query languages that have withstood the test of time and recent ones experiencing rapid adoption. The contributions of this article can be summarised as follows:
\begin{itemize}
    \item We propose eight criteria shaping what we call a \emph{Query Translation Identity Card}; each criterion represents an aspect of the translation method.
    \item We review the translation methods that exist between the most popular query languages, whereby popularity is judged based on a set of defined measures. We then categorize them based on the defined criteria.
    \item We provide a set of graphical representations of the various criteria in order to facilitate information reading, including a historical timeline of the query translation evolution.
    \item We discuss our findings, including the weakly addressed query translation paths or the unexplored ones, and report on some identified gaps and lessons learned.
\end{itemize}

\section*{Considered Query Languages}
We chose the most popular query languages in four database categories: relational, graph, hierarchical and document-oriented databases. We look at the standardization effort, number of citations to relevant publications, categorizations found in recently published works and technologies using the query languages. 
Subsequently, we introduce our chosen query languages and motivate the choice. We provide a query example for these query languages.
Our example query corresponds to the following natural language query: \textit{"Find the \underline{city} of residence of all \underline{persons} named \underline{Max}"}.

\subsection*{Relational Query Languages} 
\paragraph{SQL} is the \textit{de facto} relational query language first described in~\cite{chamberlin1974sequel}. It has been an ANSI/ISO standard since 1986/1987 and is continually receiving updates \cite{gulutzan1999sql}, latest of which was published in 2016. \\
\textbf{Example}: \texttt{SELECT place FROM Person WHERE name = "Max"}

\subsection*{Graph Query Languages} The recently published work at the ACM Computing Surveys~\cite{angles2017foundations} features three query languages: SPARQL, Cypher and Gremlin. A blogpost~\cite{topGraphDatabase} published by \textit{IBM Developer} in 2017 sees those query languages as most popular; GraphQL is also mentioned, but it has far less scientific and technological adoption.

\paragraph{SPARQL} is the \textit{de facto} language for querying RDF data. Of the three surveyed graph query languages, only SPARQL became a standard (by W3C in 2008), and is still receiving updates \cite{sparql10,sparql11}, latest of which is SPARQL 1.1 \cite{gearon2013sparql} 2013. Research articles on SPARQL foundations \cite{perez2006semantics,perez2009semantics,prudhommeaux2008sparql} are among the most cited across all graph query languages. \\
\textbf{Example}: \texttt{SELECT ?c WHERE \{?p :type :Person . ?p :name "Max" . ?p :city ?c \}}


\paragraph{Cypher} is Neo4j's query language developed in 2011, which has been open-sourced in 2015 under the OpenCypher project~\cite{green2018opencypher}. Cypher has been recently formally described in a scientific article published \cite{francis2018cypher}. At the time of writing, Neo4j tops DB engine ranking \cite{DBEnginesRanking} of Graph DBMS. \\
\textbf{Example}: \texttt{MATCH (p:Person) WHERE p.name = "Max" RETURN p.city}

\paragraph{Gremlin} \cite{rodriguez2015gremlin} is the traversal query language of Apache TinkerPop \cite{rodriguez2015gremlin}. It first appeared in 2009 and predates Cypher. It also covers wider range of graph query processing: \textit{declarative} (pattern matching) and \textit{imperative} (graph traversal). Thus, it has a larger technological adoption. For example, it has libraries in more query languages: Java, Groovy, Python, Scala, Clojure, PHP, and JavaScript; and is integrated in more renowned data processing technologies e.g., Hadoop, Spark, and graph databases, e.g., Amazon Neptune, Azure Cosmos, OrientDB, etc. \\
\textbf{Example} (\textit{declarative}): 
\texttt{g.V().match(.as('a')\\.hasLabel('Person').has('name','Max').as('p'), \_\_.as('p')\\.out('city').values().as('c')).select('c')} \\
\textbf{Example} (\textit{imperative}): 
\texttt{g.V().hasLabel('Person')\\.has('name','Max').out('city').values()}

\subsection*{Hierarchical Query Languages}
This family is dominantly represented by XML query languages.
XML appeared more than two decades ago and has been standardized in 2006 by W3C \cite{bray1997extensible}; it is used mainly for data exchange between applications. W3C recommended XML query languages are XPath and XQuery.
\paragraph{XPath} allows to define path expressions that navigate XML trees from a root parent to descendent children. XPath has been standardized by W3C in 1999, and is continually receiving updates~\cite{clark1999xml,berglund2003xml,boag2002xquery} with the latest one in 2017~\cite{dyck2017xml}.\\
\textbf{Example:} \texttt{//person[./name='Max']/city]}

\paragraph{XQuery} is the XML \textit{de facto} query language. XQuery is also considered a functional programming language, as it allows calling and writeing functions to interact with XML documents. XQuery uses XPath for path expressions, and can perform \textit{insert}, \textit{update} and \textit{delete} operations.
It has been initially suggested in 2002 \cite{boag2002xquery}, standardized by W3C in 2007 and recently updated in 2017 \cite{jonathan2017xml}.\\
\textbf{Example:} \texttt{for \$x in doc("persons.xml")/person where \$x/name='Max' return \$x/city}

\subsection*{Document Query Languages} The representative document database that we choose is \textbf{MongoDB}. MongoDB (first released in 2009) is the document database that attracted the most attention both from academia and industry. At the time of writing, MongoDB tops the DB engine ranking \cite{DBEnginesRanking} for document stores.
\paragraph{MongoDB operations.} MongoDB does not have a proper query language like SQL or SPARQL, but rather interacts with documents by means of query \textit{operations} in a JSON-like format. \\
\textbf{Example:} \texttt{db.product.find(\{name: "Max"\}, \{city: 1\})}

\section*{Query Translation Paths}
In this section, we introduce the various translation paths between the selected query languages. Figure~\ref{fig:translationPaths} shows a visual representation, where the nodes correspond to the considered query languages and the directed arrows correspond to the translation direction; the thickness of the arrows reflects the number of works on the respective query translation path.

\subsection*{SQL $<>$ XML languages}
The interest in using a relational database as a backbone for storing and querying XML has appeared as early as 1999 \cite{he1999relational}. Even though XML model differs substantially from the relation model, e.g., multi-level nesting of data, cycles, recursive graph traversals, etc., storing XML data in RDBMSs was sought to benefit from their query efficiency and storage scalability.

\paragraph{XPath/XQuery-to-SQL:} XML documents have to be \textit{flattened}, or \textit{shredded}, into relations so they can be loaded into or mapped to relational tables.
The ultimate goal is to hide the specificity of the back-end store, and make users feel as if they are directly dealing with the original XML documents.
In parallel, there are efforts to provide an XML view on top of relational databases. The rational is to unify the access, using XML, and also to benefit from XML querying capabilities, e.g., expressing path traversals and recursion.

\paragraph{SQL-to-XPath/XQuery:} This covers approaches for storing XML in native XML stores, but adding an SQL interface to enable the querying of XML by SQL users. Metadata about how XML data is mapped to the relational model is required.

\subsection*{SQL $<>$ SPARQL}
\paragraph{SPARQL-to-SQL:} Similarly to XML, the interest in bridging the gap between RDF data model and the relational model emerged as early as RDF. This was motivated by multiple and various use-cases. For example, RDBMS were suggested to store RDF data \cite{melnik2001storing,prud2004optimal}, even before SPARQL standardization. Also, the Semantic Web community suggested a well-received data integration proposal, whereby disparate relational data sources are mapped to a unified ontology model and then queried uniformly~\cite{prud2004optimal,noy2004semantic}. The concept evolved to become the popular OBDA, Ontology-Based Data Access \cite{poggi2008linking}, empowering a lot of applications today.

\paragraph{SQL-to-SPARQL:} The other direction received less attention. The main two motivations presented were enhancing interoperability between the two worlds in general, and enabling reusability of the wealth of existing relational-oriented tools over RDF data, e.g, reporting and visualization.

\subsection*{SQL-to-Document-based}
The main motivation behind exploring this path was to enable SQL users and legacy systems to access the new class of NoSQL document databases with their sole SQL knowledge.

\paragraph{SPARQL-to-Document:} The rational here is identical to that of SPARQL-to-SQL, with one extra consideration: scalability. Native triple stores become prone to scalability issues when storing and querying significant amounts of RDF data. Users resorted to more scalable solutions to store and query the data \cite{hausenblas2012large}. The most studied database solution by the research community, we found, was MongoDB.

\subsection*{SQL $<>$ Graph-based}
\paragraph{SQL-to-Cypher:} This path is considered for the same reasons as the SQL-to-Document, which is mainly attempting to help users with SQL knowledge to approach graph data stored in Neo4j.

\paragraph{Cypher-to-SQL:} The rational is to allow running graph queries over relational databases. It has also been advocated that using relational databases to store graph data can be beneficial in certain cases, benefiting from the efficient index-based retrieval RDBMSs offer.

\paragraph{Gremlin-to-SQL:} The aim here is to allow executing Gremlin traversals (without side effect steps) on top of relation databases, in order to leverage the optimization techniques built into RDBMSs. In order to do so, the property graph data is represented and stored as relational tables.

\paragraph{SQL-to-Gremlin:} The main motivation is to enable relational database users to migrate to graph databases in order to leverage the advantages of graph-based functions (e.g., depth-first search, shortest paths, etc.) and data analytical applications that require distributed graph data processing. 

\subsection*{SPARQL $<>$ XML languages}
\paragraph{SPARQL-to-XPath/XQuery:} Similarly to SQL-to-XML paths, this path seeks to build interoperability environments between semantic and XML database systems, to enable ontology-based data access to XML data, and to add a semantic layer on top XML data and services for integration purposes.

\subsection*{SPARQL $<>$ Graph-based}
\paragraph{XPath/XQuery-to-SPARQL:} Enabling XPath traversal or XQuery functional programming styles on top of RDF data can be an interesting feature to equip native RDF stores with, in order to embark adopters from the XML world into the Semantic Web world.

\paragraph{SPARQL-to-Gremlin:} This path aims to bridge the gap between the Semantic Web and Graph database communities by enabling SPARQL querying of property graph databases. Users well versed in SPARQL query language can avoid learning another query language, as Gremlin supports both OLTP and OLAP graph processors, covering a wide variety of graph databases.


\begin{figure}[!t]
    \centering
    \begin{tikzpicture}
    \tikzstyle{language}=[draw,rounded corners=4pt]
    \node[language] (sparql) at (0,-1) {SPARQL};
    \node[language] (sql) at (0,4) {SQL};
    \node[language,align=left] (xquery) at (6,2) {XPath\\XQuery};
    \node[language] (gremlin) at (9,3) {Gremlin};
    \node[language,align=left] (document) at (-5,-1) {Document\\based};
    \node[language] (cypher) at (-5,4) {Cypher};
    \draw[->,>=latex] (sql) to[bend left] node[pos=0.5,right]{\cite{ramanujam2009r2d,ramanujam2009r2dextra,rachapalli2011retro}} (sparql);
    \draw[->,>=latex] (sparql.north west) to[bend left=45] node[pos=0.2,left,text width=1.8cm,align=right]{\cite{stadlerconnecting,kiminki2010sparql,priyatna2014formalisation,Chebotko06semanticspreserving,lu2008effective,elliott2009complete,unbehauen2012accessing,sequeda2013ultrawrap,journals/ws/Rodriguez-MuroR15}} (sql.south);
    \draw[->,>=latex] (sparql.north east) to[bend right] node[pos=0.8,right,align=left]{\cite{bikakis2015sparql2xquery,bikakis2009querying,bikakis2009semantic,groppe2008embedding,fischer2011translating}} (xquery.south);
    \draw[->,>=latex] (xquery.south west) to[bend right=10] node[pos=0.7,below]{\cite{droop2007translating}} (sparql.north east) ;
    \draw[->,>=latex,rounded corners=4pt] (xquery.north) |- (sql.east) node[pos=0.68,below]{\cite{krishnamurthy2004efficient,fan2005query,mani2006join,georgiadis2007xpath,hu2008adaptive,min2008xtron}};
    \draw[->,>=latex] (sql.south east) -- (xquery.north west) node[pos=0.5,below,sloped]{\cite{vidhya2009query,vidhya2010insert,jigyasu2006sql,halverson2004rox}};
    \draw[->,>=latex,rounded corners=4pt] (sparql.east) -| (gremlin.south) node[pos=0.4,below]{\cite{thakkar2018stitch,thakkar2018two}};
    \draw[->,>=latex] (sql.north east)  to[bend left=30] node[pos=0.9,left]{\cite{sqlgremlin}} (gremlin.north west);
    \draw[->,>=latex] (gremlin.north) to[bend right=45] node[pos=0.9,above]{\cite{sun2015sqlgraph}} (sql.north);
    \draw[->,>=latex] (cypher.east) -- (sql.west) node[midway,above]{\cite{cyp2sql,steer2017cytosm}};
    \draw[->,>=latex] (sql.west) -- (document.north) node[midway,left]{\cite{querymongo,mongoDB-translator-teiid,unityjdbc}};
    \draw[->,>=latex] (sparql.west) -- (document.east) node[midway,below]{\cite{mutharaju2013d,unbehauen-semantics-2016-sparqlmap-m,botoeva2016obda,conf/ontobras/AraujoABW17,michel2016generic}};
  \end{tikzpicture}
    \caption{Query translation paths found and studied.}
    \label{fig:translationPaths}
\end{figure}
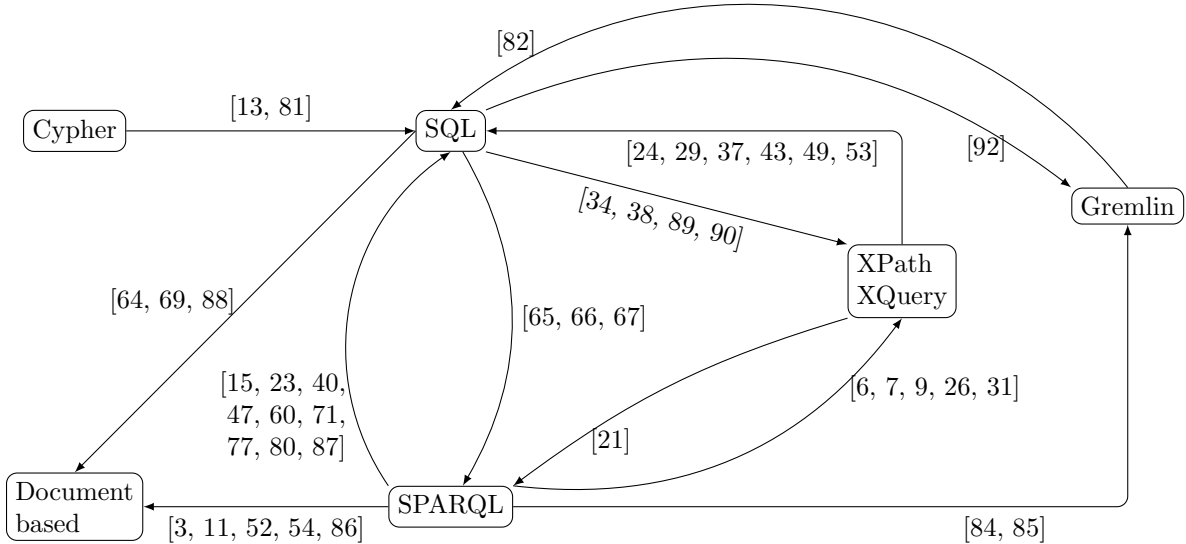


\section*{Survey Methodology}
\label{sec:TranIDCard}
Our study of the literature revealed a set of generic query translation patterns and common aspects that can be used to classify the surveyed query translation methods and tools. We refer to them as \textit{translation criteria} and organize them in three categories, forming what we call the \textit{Query Translation Identity Card}.

\subsection*{I. Translation Properties}
\begin{enumerate}[nosep]
    \item \textbf{Translation type:} Describes how the target query is obtained.
    \begin{enumerate}[nosep]
        \item \textbf{Direct:} the translation generates the destination query starting from and by analyzing \textit{only} the original query. 
        \item \textbf{Intermediate/meta query language-based:} the translation generates the destination query by passing by an intermediate (meta-)language.
        \item \textbf{Storage scheme-aware:} the translation generates queries depending on how data is internally structured or partitioned.
        \item \textbf{Schema information-aware:} the translation depends mainly on the schema information of the underlying data.
        \item \textbf{Mapping language-based:} the translation generates the destination query using a set of mapping rules expressed in an established/standardized third-party mapping language, e.g., R2RML \cite{das2011r2rml}.
    \end{enumerate}
    \item \textbf{Translation coverage:} Describes how much of the origin query language syntax is covered. For example, projection and filtering preserved, joining and update dropped.
    
    

\end{enumerate}

\subsection*{II. Translation Optimization} 
\begin{enumerate}[resume]
    \item \textbf{Optimization strategies:} Describes any optimization techniques applied during query translation, e.g., reordering joins in a query plan to reduce intermediate results.
    \item \textbf{Translation relationship:}
    Describes how many destination queries can be generated starting from the input query: one-to-one, one-to-many. Generally, it is desirable to reduce the number of destination queries to one, so we consider this an optimization aspect. We separate it from the previous point, however, as it has separate (discrete) value range.
\end{enumerate}

\subsection*{III. Community Factors}
\begin{enumerate}[resume]
    \item \textbf{Availability:} Describes whether the translation method implementation or prototype is openly available.
    That can be known, for example, by checking if the reference to the source code repository or download page is still available.
    
    \item \textbf{Adoption:} Describes the degree of acceptance of the translation method by the community by, for example, enumerating the research publications citing it. 
    
    \item \textbf{Evaluation:} Assesses whether the translation method has been empirically evaluated. For example \cite{halverson2004rox} evaluates the various schema options and their effect on query execution, using the TPC-H benchmark.
    
    \item \textbf{Metadata:} Provides some related information about the presented translation method, such as date of first and last release/update. For example, this helps to obtain an indication about whether the solution is still maintained.
\end{enumerate}

\section*{Criteria-based Classification}
\label{sec:classification}

\textbf{Scope definition.} Given the broad scope tackled in this survey, it is important to limit the search space. Therefore, we take measures as to favor quality, high-influence and completeness, as well as preserve certain level of novelty--at least in paths with the highest number of works. The measures are as follows:

\begin{itemize}[nosep]
    \item We do not consider work that describes the query translation very marginally or that has a broad scope with little focus on the query translation aspects.
    \item We only consider works proposed during the last fifteen years, i.e., after 2003. This applies in particular to XML-related translations; however, interested readers may refer to an existing survey covering older XML translation works~\cite{krishnamurthy2003xml}.
\end{itemize}

It is also important to explicitly prune the scope in terms of what is \textit{not} considered for the study:
\begin{itemize}[nosep]
    \item We do not address post-query translation steps, e.g., results format and representation.
    \item As the aim of this survey is to explore the methods and capacities, we do not comment on the results of empirical evaluations of the individual works. This is also due to the vast heterogeneity between the languages, their underlying data and use-cases.
    \item The translation method is summarized, which may entail that certain details are omitted. The goal is to allow the reader to discover the literature; interested readers are encouraged to reach to the individual publications for the full details.
\end{itemize}

In the following, we refer to the articles and tools by citation and, when applicable, by name, and directly describe the query translation methods they present. Further, it should not be inferred that the article or tool presents solely translation methods, but often, other aspects are also tackled, e.g., data migration, which are considered out-of-scope of the current study. Finally, in order to give the survey a temporal context, works are listed in a chronological order.

\subsection*{\textbf{I. Translation Properties}}
\subsubsection*{1. Translation type:}

\paragraph{(a) Direct:}

\subparagraph{SQL-to-XPath/XQuery:}
\textbf{ROX} \cite{halverson2004rox} aims at directly querying native XML stores using a SQL interface. The method consists of creating \textit{relational} views, called NICKNAMEs, over a native XML store. The NICKNAME contains schema descriptions of the rows that would be returned starting from XML input data, including mappings between those rows and XML elements expressed in form of XPath calls. Nested parent-child XML elements are caught, in the NICKNAME definition, by expressing primary and foreign keys between the corresponding NICKNAMEs. 
\cite{vidhya2010insert,vidhya2009query} propose a set of algorithms enabling direct logical translations of simple SQL INSERT, UPDATE, DELETE and RENAME queries to statements in the XUpdate language\footnote{XUpdate is an extension of XPath allowing to manipulate XML documents.}. In case of the INSERT, SQL query has to be slightly extended to instruct in which position related to the context node, preceding/following, the new node has to be inserted.

\subparagraph{SPARQL-to-SQL:}
\cite{Chebotko06semanticspreserving} 
defines a set of primitives that allow to (a) extract the relation where triples matching a triple pattern are stored, (b) extract the relational attribute whose value may match a given triple pattern in a certain position (s,p,o), (c) generate a distinct name from a triple pattern variable or URI, (d) generate SQL conditions (WHERE) given a triple pattern and the latter primitive, and (e) generate SQL projections (SELECT) given a triple pattern and the latter three primitives. A translation function returns a SQL query by fusing and building up the previous primitives given a graph pattern. The translation function generates SQL joins from UNIONs and OPTIONALs between sub-graph patters.
\textbf{FSparql2Sql} \cite{lu2008effective} is an early work focusing on the various cases of \textit{filter} in SPARQL queries. While RDF objects can take many forms like IRIs (Internationalized Resource Identifier), literals with and without language and/or datatype tags, values stored in RDBMS are generally atomic textual or numeral values. Therefore, the various cases of RDF objects are affected primitive data types, called '\textit{facets}', e.g., facets for IRIs, datatype tags and language tags are of primitive type \textit{String}. This way, filter operands become complex, so they need to be bound dynamically. To achieve that, \textit{CASE WHEN ... THEN} expressions part of SQL-92 are exploited.
\cite{elliott2009complete} proposes several translation \textit{SQL model}-algorithms implementing different operators of a SPARQL query (algebra). In contrast to many existing works, this work aims to generate \textit{flat/un-nested} SQL queries, instead of multi-level nested-queries, so SQL query optimizers can achieve better performance. This is done via SQL augmentations, i.e., SPARQL operators gradually augment the SQL query instead of creating a new nested one. The algorithms implement \textit{functions} which each generates a part of the final SQL query.

\subparagraph{SQL-to-Document-based:}
\textbf{QueryMongo} \cite{querymongo} is a Web-based translator that accepts a SQL query and generates an equivalent MongoDB query. The translation is based solely on SQL query syntax, i.e., not considering any data or schema. No explanation about the translation approach is provided.
\cite{sql-to-mongo-db-query-converter} is a library providing an API to translate SQL to MongoDB queries. The translation is based on SQL query syntax only.


\subparagraph{SPARQL-to-XPath/XQuery:}
\cite{groppe2008embedding} does not provide a direct translation of SPARQL, but SPARQL embedded inside XQuery. The method involves firstly representing SPARQL in form of \textit{tree of operators}. There are operators for projection, filtering, joining, optional and union; they declare how the output (XQuery) of the corresponding operations are represented. The translation involves data translation, from RDF to XML, and the translation of the operators to XQuery queries accordingly. An XML element with three sub-elements are created for each triple for each triple term (s, p and o). The translation from an operator into XQuery constructs is based on transformation rules, which replace the embedded SPARQL constructs with XQuery constructs. The translation from an operator into an XQuery constructs is based on transformation rules, which replace the embedded SPARQL constructs with XQuery constructs. 
In \textbf{XQL2Xquery} \cite{fischer2011translating}, variables of the basic graph patter (BGP) are mapped to XQuery values. A \textit{for} loop and a path expression is used to retrieve subjects and bind any variables encountered, then nested under every variable, iterate over the predicates and bind their variables. In a similar way, nestedly iterate over objects. Next, BGP constants and filters are mapped to XQuery \textit{where}. OPTIONAL is mapped to an XQuery \textit{function} implementing a \textit{left outer join}. For filters, XQuery value comparison are employed (e.g., eq, neq). ORDER BY is mapped to \textit{order by} in a FLWOR expression. LIMIT and OFFSET are handled using \textit{position} on the results. REDUCED is translated into a \textit{NO-OP}. 

\subparagraph{XPath/XQuery-to-SPARQL:}
\cite{droop2007translating} presents a translation method that includes data transformation from XML to RDF. During the data transformation process, XML nodes are annotated with information used to support all XPath axes. For example, type information, attributes, namespaces, parent-child relationships, information necessary for recursive XPath, etc. The above annotations conform to the structure of the generated RDF and are used to generate the final SPARQL query.

\subparagraph{Gremlin-to-SQL}
\cite{sun2015sqlgraph} propose a direct mapping approach for translating Gremlin queries (without the side effect step) to SQL queries.
The authors propose a generic technique to translate a subset of Gremlin queries (queries without side effect steps) into SQL leveraging the relational query optimizers. They propose techniques that make use of a novel schema which exploits both relational and non-relational storage for property graph data by combining relational storage with JSON storage for adjacency information and vertex and edge attributes respectively.

\subparagraph{SPARQL-to-Gremlin:}
\textbf{Gremlinator}~\cite{thakkar2018stitch,thakkar2018two} proposes a direct translation of SPARQL queries to Gremlin pattern matching traversals, by mapping each triple pattern within a SPARQL query to a corresponding single step in the Gremlin traversal language. 
This is made possible by the \texttt{match()}-step in Gremlin which offers a SPARQL-style of declarative construct. 
Within a single \texttt{match()}-step, multiple single step traversals can be combined forming a complex traversal, analogous to how multiple basic graph patterns constitute a complex SPARQL query~\cite{thakkar2017towards}.

\paragraph{(b) Intermediate/meta query language-based:}

\subparagraph{Type-ARQuE} \cite{kiminki2010sparql} uses an intermediate query language called AQL, Abstract Query Language. AQL is designed to stand between SQL and SPARQL, it extends from the relational algebra (in particular the join) and accommodates both SQL and SPARQL semantics. It is represented as a tree of expressions and joins between them, containing selects and orders. The translation process consists of three stages: (1) SPARQL query parsed and translated to AQL query, (2) AQL query undergoes a series of transformations (simplification) preparing it for SQL transformation, and (3) AQL query translated to the target SQL dialect, transforming AQL join tree to SQL join tree, along the other selects and orders expressions. Example of stage 2 simplifications: type inference, nested join flattening, join inner joins with parents, etc. 
In \cite{journals/ws/Rodriguez-MuroR15}, Datalog is used as an intermediate language between SPARQL and SQL. SPARQL query is translated into a semantics-similar Datalog program. First phase is translating SPARQL query to a set of Datalog rules. The translation adopts a syntactic variation of the method presented in \cite{polleres2013relation} by incorporating built-in predicates available in SQL and avoid negation, e.g., LeftJoin, isNull, isNotNul, NOT. Second phase is generating an SQL query starting from Datalog rules. Datalog atoms, \texttt{ans}, \texttt{triple}, \texttt{Join}, \texttt{Filter}, \texttt{LeftJoin}, are mapped to equivalent relational algebra operators. \texttt{ans} and \texttt{triple} are mapped to a projection, while \texttt{filter} and joins to equivalent relational filter and joins, respectively. 

\subparagraph{SPARQL-to-Document:}
In \cite{michel2016generic} a generic two-step SPARQL-to-X approach is suggested, with a showcase using MongoDB. The article proposes to convert a SPARQL query to a pivot intermediate query language called Abstract Query Language (AQL). The translation uses a set of mappings in xR2RML mapping language, which describe how data in target databases are mapped into RDF model, without converting data to RDF. AQL has a grammar that is similar to SQL both syntactically and semantically. 
The BGP part of a SPARQL query, is decomposed into a set of expressions in AQL. Next, xR2RML mappings are checked for any maps matching the containing triple patterns. Those detected matching maps are used to translate individual triple patterns to \textit{atomic abstract queries}.
Queries in AQL are translated to the query language of the target database. Unsupported operations like JOIN in MongoDB are assumed left to a higher-lever query engine. 

\paragraph{(c) Storage scheme-aware:} 

\subparagraph{XPath/XQuery-to-SQL:} In \cite{min2008xtron} \textbf{XTRON}, a relational XML management system is presented. The article suggests a schema-oblivious way of storing and querying XML data. XML documents are stored uniformly in \textit{identical} relational tables using a fixed predefined relational model. Generated queries then have to abide by this fixed relational schema scheme.

\subparagraph{SPARQL-to-Document:}
\textbf{D-SPARQ} \cite{mutharaju2013d} focuses on the efficient processing of join operation between triple patterns of a SPARQL query. RDF data is physically materialized in a cluster of MongoDB stores, following a specific graph partitioning scheme. SPARQL queries are converted to MongoDB queries following the same.

\subparagraph{Cypher-to-SQL:}
\textbf{Cyp2sql} \cite{cyp2sql} is a tool for the automatic transformation of both data and queries from Neo4j to a relational database. During the transformation, the following tables are created: Nodes, Edges, Labels, Relationship types, plus materialized views to store the adjacency list of the nodes. Cypher queries are then translated to SQL queries tailored to that data storage scheme. 

\subparagraph{SQL-to-Gremlin:} SQL-Gremlin \cite{sqlgremlin} is a proof-of-concept SQL-to-Gremlin translator. The translation requires that the underlying graph data is given a relational schema, where elements from the graph are mapped to tables and attributes. However, there is no reported scientific study that discusses the translation approach.
\textbf{SQL2Gremlin} \cite{sql2gremlin} is a tool for converting SQL queries to Gremlin queries. They show how to reproduce the effect of SQL queries using Gremlin traversals. A pre-defined graph model is used during the translation; as an example, Northwind relational data was loaded as a graph inside Gremlin.

\paragraph{(d) Schema information-aware:} 

\subparagraph{XPath/XQuery-to-SQL:} \cite{krishnamurthy2004efficient} The process uses summary information on the relational integrity constraints pre-computed in a pre-processing phase. An XML view is constructed by mapping elements from the XML schema to elements from the relational schema. The XML view is a tree where the nodes map to table names and the leaves to column names. An SQL query is built by going from the root to the leaves of this tree, a traversal from a node to a node is a join between the two corresponding tables. 
In~\cite{fan2005query} XML data is shredded into relations based on an XML schema (DTD) and saved in a RDBMS. The article extends XPath expressions to allow capturing recursive queries against a recursive schema. XPath queries with the extended expressions can, next, be translated into an equivalent sequence of SQL queries using a common RDBMS operator (LFP: Simple Least Fixpoint).
Whereas~\cite{mani2006join} builds a virtual XML view on top of relational databases using XQuery, the focus of the article is on the optimization of the intermediate relational algebra.

\subparagraph{SQL-to-SPARQL:}
\textbf{R2D} \cite{ramanujam2009r2d,ramanujam2009r2dextra} propose to create a relational virtual normalized schema (view) on top of RDF data.
Schema elements are extracted from RDF schema; if schema is missing or incomplete, schema information is extracted by thoroughly exploring the data. \textit{r2d:TableMap, r2d:keyField, r2d:refersToTableMap} denote a relational table, its primary key, and foreign key, respectively.
A relational view is created using those schema constructs, against which SQL queries are posed.
SQL queries are translated into SPARQL queries.
For every SQL projected, filtered or aggregated (with GROUP BY) variable, a variable is added to SPARQL SELECT. SQL WHERE conditions are added to SPARQL FILTER, LIKE mapped to a regex(), and blank nodes are used in a number of cases.
In \textbf{RETRO} \cite{rachapalli2011retro}
RDF data is exhaustively parsed to extract domain-specific relational schema. The schema corresponds to the so-called vertical partitioning, i.e., one table for every extracted predicate, each table is composed of $<$subject object$>$ attributes. Then, the translation algorithm parses the SQL query posed against the extracted relational schema and iteratively builds the SPARQL query. 

\subparagraph{SQL-to-Document-based: }
\cite{mongoDB-translator-teiid} requires the user to provide a MongoDB schema, expressed in a relational form using tables, procedures, and functions. 
\cite{unityjdbc} provides a JDBC access to MongoDB documents by building a representative schema, which is, in turn, constructed by sampling MongoDB data and fitting the least-general type representing the data.

\subparagraph{SQL-to-XPath/XQuery:}
\textbf{AquaLogic Data Services Platform} \cite{jigyasu2006sql} builds an XML-based layer on top of heterogeneous data sources and services. To allow SQL access to relational data, relational schema is mapped to AquaLogic DSP artifacts (internal data organization), e.g., service function to relational tables.  

\subparagraph{SPARQL-to-Document:}
\cite{botoeva2016obda}, in the context of OBDA, suggests a two-step approach, whereby the relational model is used as an intermediate model between SPARQL and MongoDB queries. Notions of MongoDB \textit{type constrains} (schema) and \textit{mapping assertions} are imposed on MongoDB data, both of which are used during the first phase of query translation to create relational views. The schema is extracted from the data stored in MongoDB. MongoDB mappings relate MongoDB paths (e.g., student.name) to ontology properties. A SPARQL query is first decomposed into a set of \textit{translatable} sub-queries. Using MongoDB mappings, MongoDB queries are created. 
\textbf{OntoMongo} \cite{conf/ontobras/AraujoABW17} proposes an OBDA on top of NoSQL stores, applied to MongoDB. An ontology, conceptual layer, and mapping between the ontology and conceptual layer are involved. The conceptual layer adopts the object-oriented programming model, i.e., classes and hierarchy of classes. Data is accessed via ODM, Document-Relational Mapping, calls. SPARQL triple patterns are grouped by their shared subject variable (star-shaped). Each group of triples is assumed to be of one class defined in the mappings, the class name is denoted by the variable of the shared subject. MongoDB query can be created by mapping query classes to classes in the conceptual model, which then is used to call MongoDB terms via the ODM. The lack of JOIN operation in MongoDB is substituted with a combination of two unwind commands each concerning one side (class) of the join. 

\subparagraph{Cypher-to-SQL:}
\textbf{Cytosm} \cite{steer2017cytosm} presents a middleware allowing to execute graph queries directly on non-graph databases. The application relies on gTop (graph Topology) to build a form of schema on top of graph data. gTop consists of two components: (1) Abstract Property Graph model and (2) a mapping to the relational model. It captures the structure of \textit{property} graphs, i.e., node and edge types and their properties, and provides mapping between graph query language and the relational query language, mapping nodes to rows of tables, and edges to either fields of rows or a sequence of table-join operations.
Query translation is twofold. (1) Using gTop abstract model, Cypher path expressions (from MATCH keyword) are visited and a set of \textit{restricted} OpenCypher~\cite{francis2018cypher} queries not containing multi-hop edges and anonymous entities (which are not possible to translate to SQL) are generated, denoted rOCQ. (2) rOCQ are parsed and an intermediate SQL-like representation is generated, having one SELECT and WITH SELECT for each MATCH. SELECT variables are checked if they require information from the RDBMS, and if they inter-depend. Then, the mapping part of gTop is used to map nodes to relational tables. Finally, edges are resolved into JOINs, also basing on gTop mappings. 

\subparagraph{SPARQL-to-XPath/XQuery:}
\textbf{SPARQL2XQuery} is described in a couple of publications \cite{bikakis2015sparql2xquery,bikakis2009querying,bikakis2009semantic}. The translation is based on a mapping model between OWL ontology (existing or user-defined) and XML Schema. Mappings can either be automatically extracted by analyzing the ontology and XML schema, or manually curated by a domain expert. SPARQL queries are posed against the ontology without knowledge of the XML schema. The BGP (Basic Graph Pattern) of SPARQL query is normalized into a form where each GP is UNION-free, so each GP can be processed independently and more efficiently. XPaths are bound to GP variables, there are various forms of binding for various types of variables. Next, GPs are translated into an equivalent XQuery expression using the mappings; for each variable of a triple, a \textit{For} or \textit{Let} clause using the variable binding is created. 
\textbf{Ultrawrap} \cite{sequeda2013ultrawrap} implements an RDF2RDB mapping, allowing to execute SPARQL queries on top of existing RDBMSs. It creates an RDF ontology from the SQL schema, based on which it next creates a set of logical RDF views over the RDBMS. The views, called \textit{Tripleviews}, are an extension of the famous \textit{triple tables} (subject,predicate,object) with two additional columns: subject and object primary keys. Four Tripleviews are created: \textit{types} ---stores subjects along their types in the DB, \textit{varchar(size)} ---stores only textual attributes, \textit{int} ---stores only numeral attributes, and \textit{object properties} ---stores join links between DB tables. Given a SPARQL query, each triple pattern maps to a Tripleview. 

\paragraph{(e) Mapping language-based:}

\subparagraph{SPARQL-to-SQL:}
In \textbf{SparqlMap} \cite{unbehauen2012accessing} triple patterns of a SPARQL query are individually examined to extract R2RML triple maps. Methods are applied to find the candidate set of triple maps, and then to prune this to produce a set that prepares for the subsequent query translation. Given a SPARQL query, a recursive query generation process is devised yielding a \textit{single} but nested SQL query. Sub-queries are created for individual mapped triple patterns and for reconciling those via JOIN or UNION. Nested subqueries querying the RDBMS tables extract not only the columns but also structural information like term type (resource, literal, etc.), concatenates multiple columns to form IRIs, etc. To generalize the technique of \cite{journals/ws/Rodriguez-MuroR15} (Datalog as intermediate language) to arbitrary relational schema, R2RML is incorporated. For every R2RML \textit{triple map} a set of Datalog rules are generated reflecting the same semantics. A \texttt{triple} atom is created for every combination of \textit{subject map}, \textit{property map} and \textit{object map} on a translated \textit{logical table}. Finally, the translation process from Datalog to SQL is extended to deal with the new rules introduced  by R2RML mappings. 
\cite{priyatna2014formalisation} extends a previously published translation method \cite{Chebotko06semanticspreserving} to involve user-defined R2RML mappings. 
In particular, it incorporates R2RML mappings in $\alpha$ and $\beta$ mappings as well as \textit{genCondSQL()}, \textit{genPRSQL()} and \textit{trans()} functions. 
For each, an algorithm is devised, considering the various situations found in R2RML mappings like the absence of Reference Object Map. 
\textbf{SparqlMap-M} \cite{unbehauen-semantics-2016-sparqlmap-m} enables querying document stores using SPARQL without RDF data materialization. It is based on a previous SPARQL-to-SQL translator, SparqlMap \cite{unbehauen2012accessing}, so it adopts a relational model to virtually represent the data. Documents are mapped to relations using an extension of R2RML allowing to capture duplicate demoralized data, which is common characteristic of document data. The lack of union and join capabilities support is mitigated by a multi-level query execution, producing and reusing intermediate results. Selection parts are pushed to the document store, while the union and join are executed using an internal RDF store.

\subsubsection*{2. Translation coverage:}

We note the following before starting our review of the works:
\begin{itemize}[nosep]
    \item The coverage is extracted not only from the core of the articles, but also from the evaluation sections and from the online page of the implementations (when available). For example, \cite{sequeda2013ultrawrap,unbehauen-semantics-2016-sparqlmap-m} evaluate using all 12 BSBM benchmark queries, which cover more scope than that of the article; the corresponding Web page of \cite{sequeda2013ultrawrap} mention features that are both beyond the core and the evaluation section of the article.
    \item We mention the supported query feature but we do not assume its completeness, e.g., \cite{conf/ontobras/AraujoABW17} supports filters but only for \textit{equality} condition. Interested users are encouraged to seek details from the corresponding articles/tools.
    \item Table~\ref{table:SPARQLCoverage} shows that some works \cite{Chebotko06semanticspreserving,conf/ontobras/AraujoABW17} support only one feature. This does not necessarily imply insignificance, but reflects a choice to reserve the full study to covering that particular feature, e.g., various shapes of graph patters or different cases of OPTIONAL.
\end{itemize}
\subparagraph{SQL-to-X and SPARQL-to-X:} See Table \ref{table:SQLCoverage} and Table~\ref{table:SPARQLCoverage} for translation methods and tools from SQL to SPARQL respectively. For SQL, the \texttt{WHERE} clause is an essential part of most useful queries, hence, it is supported by all methods. \texttt{GROUP BY} is the next commonly supported feature, as it enables a significant class of SQL queries: analytical and aggregational queries. To a lower extent supported is the sorting operation \texttt{ORDER BY}. \texttt{UNION} and especially \texttt{JOIN} are operations of typically high cost; they are among the least supported features.
As most researched query categories are of retrieval nature, modification queries such as \texttt{INSERT}, \texttt{UPDATE} and \texttt{DELETE} are very weakly addressed.
\texttt{DISTINCT} and nested queries are rarely supported, which might also be attributed to their typical expensiveness, e.g., \texttt{DISTINCT} requires sorting, and nested-queries generate large intermediate results. \texttt{EXCEPT}, \texttt{UPSERT}, and \texttt{CREATE} are only supported by individual works.
For SPARQL, query operation support is more prominent across the reviewed works. \texttt{FILTER}, \texttt{UNION} and \texttt{OPTIONAL} are the most commonly supported query operations with up to 60\% of the surveyed works. To less extent, 
\texttt{DISTINCT}, \texttt{LIMIT} and \texttt{ORDER BY} are supported by about half of the works. The rest query operations are all supported by a few works , e.g., \texttt{DESCRIBE}, \texttt{CONSTRUCT}, \texttt{ASK}, blank nodes, \texttt{datatype()}, \texttt{bound()}, \texttt{isLiteral()}, \texttt{isURI()}, etc. \texttt{GRAPH}, \texttt{SUB-GRAPH}, \texttt{BIND} are examples of interesting query operations but only supported by individual works. In general, \texttt{DESCRIBE}, \texttt{CONSTRUCT} and \texttt{ASK} are far less prominent SPARQL query constructs in comparison to \texttt{SELECT}, which is present in all the works. \texttt{isURI()} and \texttt{isLiteral()} are SPARQL-specific functions with no direct equivalent in other languages.

\subparagraph{XPath/XQuery-to-SQL:}
The queries \cite{krishnamurthy2004efficient} focuses on are simple path expressions, including descendent axis traversal, i.e., //.
\cite{fan2005query} enables XPath recursive queries against a recursive schema.
\cite{mani2006join} focuses on optimizing relational algebra, only a simple XPath query is used for the example.
\cite{georgiadis2007xpath} covers simple, ancestor, following, parent, following-sibling, descendant-or-self XPath queries.
In \cite{hu2008adaptive}, the supported queries are XPath queries with descendent/child axes with simple conditions. 
\cite{min2008xtron} translates XQuery queries with path expressions including decedent axis // XQuery queries, dereference operator $=>$ and FLWR expressions.

\subparagraph{XPath/XQuery-to-SPARQL:}
\cite{droop2007translating} mentions support for recursive XPath queries, with descendent, following and preceding axes as well as for filters.

\subparagraph{Cypher-to-SQL:}
\cite{steer2017cytosm} experiments with queries containing \texttt{MATCH}, \texttt{WITH}, \texttt{WHERE}, \texttt{RETURN}, \texttt{DISTINCT}, \texttt{CASE}, \texttt{ORDER BY}, \texttt{LIMIT}, and with patters: simple patterns with known nodes and relationships, and $->$ and $<-$ directions, variable-length relationship.
\cite{cyp2sql} is able to translate \texttt{MATCH}, \texttt{WITH}, \texttt{WHERE}, \texttt{RETURN}, \texttt{DISTINCT}, \texttt{ORDER BY}, \texttt{LIMIT}, \texttt{SKIP}, \texttt{UNION}, count(), collect(), exists(), label(), id(), and rich pattern cases, e.g., (a or empty)--()--(b or empty), [a or empty]-[b]-(c or empty), $->$ and $<-$, (a) $-->$ (b).

\begin{table}[p]
\centering
\rotatebox{90}{
\begin{tabular}{|>{\small}l|>{\scriptsize}c|>{\scriptsize}c|>{\scriptsize}c|>{\scriptsize}c|>{\scriptsize}c|>{\scriptsize}c|>{\scriptsize}c|>{\scriptsize}c|>{\scriptsize}c|>{\scriptsize}c|>{\scriptsize}p{.12\columnwidth}|}
\hline
\rowcolor{blue!35}\textbf{Work} & \textbf{\texttt{DISTINCT}} & \begin{tabular}[c]{@{}l@{}}\textbf{\texttt{WHERE}}/\\\textbf{\texttt{REGEX}}\end{tabular} & \textbf{\texttt{JOIN}} & \begin{tabular}[c]{@{}l@{}}\textbf{\texttt{UNION}}\end{tabular} & \begin{tabular}[c]{@{}l@{}}\textbf{\texttt{GROUP BY}}\\ /\textbf{\texttt{HAVING}}\end{tabular} & \textbf{\texttt{ORDER BY}} & \begin{tabular}[c]{@{}l@{}}\textbf{\texttt{LIMIT}}/\\ \textbf{\texttt{OFFSET}}\end{tabular} &  \begin{tabular}[c]{@{}l@{}}\textbf{\texttt{INSERT}}/\\ \textbf{\texttt{UPDATE}}\end{tabular} &
\begin{tabular}[c]{@{}l@{}}\textbf{\texttt{DELETE}}/\\ \textbf{\texttt{DROP}}\end{tabular} &
\begin{tabular}[c]{@{}l@{}}\textbf{Nested}\\ \textbf{queries}\end{tabular} & \textbf{Others} \\ \hline
\rowcolor{LightCyan}\multicolumn{12}{c}{\textit{SQL-to-XPath/XQuery}} \\ \hline
\cite{halverson2004rox} & {\Large \textcolor{orange}{\textbf{?}}} & \textcolor{checked}{\CheckmarkBold}/ & {\Large \textcolor{orange}{\textbf{?}}} & {\Large \textcolor{orange}{\textbf{?}}} & \textcolor{checked}{\CheckmarkBold}/ & \textcolor{checked}{\CheckmarkBold}  & {\Large \textcolor{orange}{\textbf{?}}} & {\Large \textcolor{orange}{\textbf{?}}} & \textcolor{checked}{\CheckmarkBold}/ & {\Large \textcolor{orange}{\textbf{?}}} & \\ \hline
\cite{jigyasu2006sql} & {\Large \textcolor{orange}{\textbf{?}}} & \textcolor{checked}{\CheckmarkBold}/ & \textcolor{checked}{\CheckmarkBold} & \textcolor{checked}{\CheckmarkBold} & {\Large \textcolor{orange}{\textbf{?}}} & \textcolor{checked}{\CheckmarkBold} & {\Large \textcolor{orange}{\textbf{?}}} & {\Large \textcolor{orange}{\textbf{?}}} & {\Large \textcolor{orange}{\textbf{?}}} & \textcolor{checked}{\CheckmarkBold} & \\ \hline
\cite{vidhya2009query,vidhya2010insert} & {\Large \textcolor{orange}{\textbf{?}}} & \textcolor{checked}{\CheckmarkBold}/ & {\Large \textcolor{orange}{\textbf{?}}} & {\Large \textcolor{orange}{\textbf{?}}} & {\Large \textcolor{orange}{\textbf{?}}} & {\Large \textcolor{orange}{\textbf{?}}} & {\Large \textcolor{orange}{\textbf{?}}} & \textcolor{checked}{\CheckmarkBold}/\textcolor{checked}{\CheckmarkBold} & \textcolor{checked}{\CheckmarkBold}/ & {\Large \textcolor{orange}{\textbf{?}}} & \texttt{RENAME} \\ \hline
\rowcolor{LightCyan}\multicolumn{12}{c}{\textit{SQL-to-SPARQL}} \\ \hline
\cite{rachapalli2011retro} & {\Large \textcolor{orange}{\textbf{?}}} & \textcolor{checked}{\CheckmarkBold}/ & \textcolor{checked}{\CheckmarkBold} & \textcolor{checked}{\CheckmarkBold} &  & {\Large \textcolor{orange}{\textbf{?}}} & {\Large \textcolor{orange}{\textbf{?}}} & {\Large \textcolor{orange}{\textbf{?}}} & {\Large \textcolor{orange}{\textbf{?}}} & {\Large \textcolor{orange}{\textbf{?}}} & \texttt{EXCEPT} \\ \hline
\cite{ramanujam2009r2d,ramanujam2009r2dextra} & {\Large \textcolor{orange}{\textbf{?}}} & \textcolor{checked}{\CheckmarkBold}/\textcolor{checked}{\CheckmarkBold} & {\Large \textcolor{orange}{\textbf{?}}} & {\Large \textcolor{orange}{\textbf{?}}} & \textcolor{checked}{\CheckmarkBold}/ & {\Large \textcolor{orange}{\textbf{?}}} & {\Large \textcolor{orange}{\textbf{?}}} & {\Large \textcolor{orange}{\textbf{?}}} & {\Large \textcolor{orange}{\textbf{?}}} & {\Large \textcolor{orange}{\textbf{?}}} & \\ \hline
\rowcolor{LightCyan}\multicolumn{12}{c}{\textit{SQL-to-Document-based}} \\ \hline
\cite{querymongo} & \textcolor{checked}{\CheckmarkBold} & \textcolor{checked}{\CheckmarkBold}/\textcolor{checked}{\CheckmarkBold} & \textcolor{red}{\XSolidBrush} & \textcolor{red}{\XSolidBrush} & \textcolor{checked}{\CheckmarkBold}/\textcolor{checked}{\CheckmarkBold} & \textcolor{checked}{\CheckmarkBold} & \textcolor{checked}{\CheckmarkBold}/ & \textcolor{red}{\XSolidBrush} & \textcolor{red}{\XSolidBrush} & \textcolor{red}{\XSolidBrush} & \\ \hline
\cite{mongoDB-translator-teiid} & \textcolor{checked}{\CheckmarkBold} & \textcolor{checked}{\CheckmarkBold}/ & {\Large \textcolor{orange}{\textbf{?}}} & {\Large \textcolor{orange}{\textbf{?}}} & \textcolor{checked}{\CheckmarkBold}/\textcolor{checked}{\CheckmarkBold} & \textcolor{checked}{\CheckmarkBold} & \textcolor{checked}{\CheckmarkBold}/\textcolor{checked}{\CheckmarkBold} & \textcolor{checked}{\CheckmarkBold}/ & \textcolor{checked}{\CheckmarkBold}/ & {\Large \textcolor{orange}{\textbf{?}}} & \\ \hline
\cite{unityjdbc} & {\Large \textcolor{orange}{\textbf{?}}} & \textcolor{checked}{\CheckmarkBold}/\textcolor{checked}{\CheckmarkBold} & \textcolor{checked}{\CheckmarkBold} & {\Large \textcolor{orange}{\textbf{?}}} & \textcolor{checked}{\CheckmarkBold}/ & {\Large \textcolor{orange}{\textbf{?}}} & \textcolor{checked}{\CheckmarkBold}/\textcolor{checked}{\CheckmarkBold} & \textcolor{checked}{\CheckmarkBold}/\textcolor{checked}{\CheckmarkBold} & /\textcolor{checked}{\CheckmarkBold} & {\Large \textcolor{orange}{\textbf{?}}} & \texttt{CREATE}, \texttt{DROP}, \texttt{UPSERT}, \textit{date, string, math fncts} \\ \hline
\cite{sql-to-mongo-db-query-converter} & {\Large \textcolor{orange}{\textbf{?}}} & \textcolor{checked}{\CheckmarkBold}/\textcolor{checked}{\CheckmarkBold} & {\Large \textcolor{orange}{\textbf{?}}} & {\Large \textcolor{orange}{\textbf{?}}} & \textcolor{checked}{\CheckmarkBold}/ & \textcolor{checked}{\CheckmarkBold} & {\Large \textcolor{orange}{\textbf{?}}} & {\Large \textcolor{orange}{\textbf{?}}} & \textcolor{checked}{\CheckmarkBold}/ & {\Large \textcolor{orange}{\textbf{?}}} & some Boolean filters \\ \hline
\cite{neo4jsql} & {\Large \textcolor{orange}{\textbf{?}}} & \textcolor{checked}{\CheckmarkBold}/ & \textcolor{checked}{\CheckmarkBold} & {\Large \textcolor{orange}{\textbf{?}}} & \textcolor{checked}{\CheckmarkBold}/ & \textcolor{checked}{\CheckmarkBold} & {\Large \textcolor{orange}{\textbf{?}}} & {\Large \textcolor{orange}{\textbf{?}}} & {\Large \textcolor{orange}{\textbf{?}}} & {\Large \textcolor{orange}{\textbf{?}}} & \\ \hline
\rowcolor{LightCyan}\multicolumn{12}{c}{\textit{SQL-to-Gremlin}} \\ \hline
\cite{sqlgremlin} & \textcolor{checked}{\CheckmarkBold} & \textcolor{checked}{\CheckmarkBold}/\textcolor{checked}{\CheckmarkBold} & {\Large \textcolor{orange}{\textbf{?}}} & \textcolor{checked}{\CheckmarkBold} & \textcolor{checked}{\CheckmarkBold}/ & \textcolor{checked}{\CheckmarkBold} & {\Large \textcolor{orange}{\textbf{?}}} & {\Large \textcolor{orange}{\textbf{?}}} & {\Large \textcolor{orange}{\textbf{?}}} & \textcolor{checked}{\CheckmarkBold} & \\ \hline
\end{tabular}
}

\caption{SQL features supported in SQL-to-\textit{X} query translations. \textcolor{checked}{\CheckmarkBold}~is supported, \textcolor{red}{\XSolidBrush}~is not supported, {\Large \textcolor{orange}{\textbf{?}}} not (clearly) mentioned supported. \textit{Others} are features provided only by individual works.} \vspace{-10pt}
\label{table:SQLCoverage}
\end{table}

\begin{table}[p]
\centering
\small
\rotatebox{90}{
\begin{tabular}{|>{\scriptsize}l|>{\scriptsize}c|>{\scriptsize}c|>{\scriptsize}c|>{\scriptsize}c|>{\scriptsize}c|>{\scriptsize}c|>{\scriptsize}c|>{\scriptsize}c|>{\scriptsize}c|>{\scriptsize}c|>{\scriptsize}c|>{\scriptsize}p{.05\columnwidth}|}
\hline
\rowcolor{blue!35}\textbf{Work} & \begin{tabular}[c]{@{}l@{}}\textbf{\texttt{DISTINCT}}\\/\textbf{\texttt{REDUCED}}\end{tabular} & 
\begin{tabular}[c]{@{}l@{}}\textbf{\texttt{FILTER}}/\\\textbf{\texttt{regex()}}\end{tabular} & \textbf{\texttt{OPTIONAL}} & \begin{tabular}[c]{@{}l@{}}\textbf{\texttt{UNION}}\end{tabular} & \textbf{\texttt{ORDER BY}} & \begin{tabular}[c]{@{}l@{}}\textbf{\texttt{LIMIT}}/\\ \textbf{\texttt{OFFSET}}\end{tabular} & 
\begin{tabular}[c]{@{}l@{}}\textbf{Blank}\\ \textbf{nodes}\end{tabular} & 
\begin{tabular}[c]{@{}l@{}}\textbf{\texttt{datatype()}}\\ \textbf{\texttt{/lang()}}\end{tabular} &
\begin{tabular}[c]{@{}l@{}}\textbf{\texttt{isURI()}}\\ \textbf{\texttt{isLiteral()}}\end{tabular} &
\begin{tabular}[c]{@{}l@{}}\textbf{\texttt{DESCRIBE}}\\ /\textbf{\texttt{bound()}}\end{tabular} &
\begin{tabular}[c]{@{}l@{}}\textbf{\texttt{CONSTRUCT}}\\ /\textbf{\texttt{ASK}}\end{tabular} &
\textbf{Others} \\ \hline
\rowcolor{LightCyan}\multicolumn{13}{c}{\textit{SPARQL-to-SQL}} \\ \hline
\cite{kiminki2010sparql} & \textcolor{checked}{\CheckmarkBold}/ & {\Large \textcolor{orange}{\textbf{?}}} & {\Large \textcolor{orange}{\textbf{?}}} & \textcolor{checked}{\CheckmarkBold} & \textcolor{red}{\XSolidBrush} & \textcolor{checked}{\CheckmarkBold}/ & {\Large \textcolor{orange}{\textbf{?}}} & {\Large \textcolor{orange}{\textbf{?}}} & {\Large \textcolor{orange}{\textbf{?}}} & {\Large \textcolor{orange}{\textbf{?}}} & {\Large \textcolor{orange}{\textbf{?}}} & \\ \hline
\cite{priyatna2014formalisation} & {\Large \textcolor{orange}{\textbf{?}}} & {\Large \textcolor{orange}{\textbf{?}}} & {\Large \textcolor{orange}{\textbf{?}}} & {\Large \textcolor{orange}{\textbf{?}}} & {\Large \textcolor{orange}{\textbf{?}}} & {\Large \textcolor{orange}{\textbf{?}}} & {\Large \textcolor{orange}{\textbf{?}}} & {\Large \textcolor{orange}{\textbf{?}}} & {\Large \textcolor{orange}{\textbf{?}}} & {\Large \textcolor{orange}{\textbf{?}}} & {\Large \textcolor{orange}{\textbf{?}}} & \\ \hline
\cite{elliott2009complete} & \textcolor{checked}{\CheckmarkBold}/ & {\Large \textcolor{orange}{\textbf{?}}} & \textcolor{checked}{\CheckmarkBold} & \textcolor{checked}{\CheckmarkBold} & \textcolor{checked}{\CheckmarkBold} & \textcolor{checked}{\CheckmarkBold}/ & \textcolor{checked}{\CheckmarkBold} & \textcolor{checked}{\CheckmarkBold}/ & \textcolor{checked}{\CheckmarkBold} & \textcolor{checked}{\CheckmarkBold}/ & \textcolor{checked}{\CheckmarkBold}/\textcolor{checked}{\CheckmarkBold} & \texttt{GRAPH}, \texttt{FROM NAMED}, \textit{isBlank()} \\ \hline
\cite{unbehauen2012accessing} & {\Large \textcolor{orange}{\textbf{?}}}/ & {\Large \textcolor{orange}{\textbf{?}}} & \textcolor{checked}{\CheckmarkBold} & {\Large \textcolor{orange}{\textbf{?}}} & \textcolor{checked}{\CheckmarkBold} & {\Large \textcolor{orange}{\textbf{?}}} & {\Large \textcolor{orange}{\textbf{?}}} & {\Large \textcolor{orange}{\textbf{?}}} & {\Large \textcolor{orange}{\textbf{?}}} & {\Large \textcolor{orange}{\textbf{?}}} & {\Large \textcolor{orange}{\textbf{?}}} & \\ \hline
\cite{thakkar2018stitch,thakkar2018two} & \textcolor{checked}{\CheckmarkBold}/\textcolor{red}{\XSolidBrush} & \textcolor{checked}{\CheckmarkBold}/\textcolor{checked}{\CheckmarkBold} & \textcolor{checked}{\CheckmarkBold} & \textcolor{checked}{\CheckmarkBold} & \textcolor{checked}{\CheckmarkBold} & \textcolor{checked}{\CheckmarkBold}/\textcolor{checked}{\CheckmarkBold} & \textcolor{checked}{\CheckmarkBold} & {\Large \textcolor{orange}{\textbf{?}}} & {\Large \textcolor{orange}{\textbf{?}}} & \textcolor{red}{\XSolidBrush}/\textcolor{red}{\XSolidBrush} & \textcolor{red}{\XSolidBrush}/\textcolor{red}{\XSolidBrush} & \texttt{GROUP BY}, SUBGRAPH, \texttt{REMOTE} \\ \hline
\cite{lu2008effective} & {\Large \textcolor{orange}{\textbf{?}}} & \textcolor{checked}{\CheckmarkBold} & \textcolor{checked}{\CheckmarkBold} & \textcolor{checked}{\CheckmarkBold} & {\Large \textcolor{orange}{\textbf{?}}} & {\Large \textcolor{orange}{\textbf{?}}} & {\Large \textcolor{orange}{\textbf{?}}} & \textcolor{checked}{\CheckmarkBold}/ & /\textcolor{checked}{\CheckmarkBold} & /\textcolor{checked}{\CheckmarkBold} & {\Large \textcolor{orange}{\textbf{?}}} & \\ \hline
\cite{sequeda2013ultrawrap} & \textcolor{checked}{\CheckmarkBold}/ & \textcolor{checked}{\CheckmarkBold}/\textcolor{checked}{\CheckmarkBold} & \textcolor{checked}{\CheckmarkBold} & \textcolor{checked}{\CheckmarkBold} & \textcolor{checked}{\CheckmarkBold} & \textcolor{checked}{\CheckmarkBold}/\textcolor{checked}{\CheckmarkBold} & {\Large \textcolor{orange}{\textbf{?}}} & /\textcolor{checked}{\CheckmarkBold} & {\Large \textcolor{orange}{\textbf{?}}} & \textcolor{checked}{\CheckmarkBold}/\textcolor{checked}{\CheckmarkBold} & {\Large \textcolor{orange}{\textbf{?}}} & \texttt{BIND} \\ \hline
\cite{journals/ws/Rodriguez-MuroR15} & \textcolor{checked}{\CheckmarkBold}/ & \textcolor{checked}{\CheckmarkBold}/ & \textcolor{checked}{\CheckmarkBold} & \textcolor{checked}{\CheckmarkBold} & \textcolor{checked}{\CheckmarkBold} & \textcolor{checked}{\CheckmarkBold}/\textcolor{checked}{\CheckmarkBold} & {\Large \textcolor{orange}{\textbf{?}}} & /\textcolor{checked}{\CheckmarkBold} & {\Large \textcolor{orange}{\textbf{?}}} & {\Large \textcolor{orange}{\textbf{?}}} & {\Large \textcolor{orange}{\textbf{?}}} & \\ \hline
\cite{Chebotko06semanticspreserving} & {\Large \textcolor{orange}{\textbf{?}}} & {\Large \textcolor{orange}{\textbf{?}}} & \textcolor{checked}{\CheckmarkBold} & {\Large \textcolor{orange}{\textbf{?}}} & {\Large \textcolor{orange}{\textbf{?}}} & {\Large \textcolor{orange}{\textbf{?}}} & {\Large \textcolor{orange}{\textbf{?}}} & {\Large \textcolor{orange}{\textbf{?}}} & {\Large \textcolor{orange}{\textbf{?}}} & {\Large \textcolor{orange}{\textbf{?}}} & {\Large \textcolor{orange}{\textbf{?}}} & \\ \hline
\rowcolor{LightCyan}\multicolumn{13}{c}{\textit{SPARQL-to-Document}} \\ \hline
\cite{conf/ontobras/AraujoABW17} & {\Large \textcolor{orange}{\textbf{?}}} & \textcolor{checked}{\CheckmarkBold}/ & {\Large \textcolor{orange}{\textbf{?}}} & {\Large \textcolor{orange}{\textbf{?}}} & {\Large \textcolor{orange}{\textbf{?}}} & {\Large \textcolor{orange}{\textbf{?}}} & {\Large \textcolor{orange}{\textbf{?}}} & {\Large \textcolor{orange}{\textbf{?}}} & {\Large \textcolor{orange}{\textbf{?}}} & {\Large \textcolor{orange}{\textbf{?}}} & {\Large \textcolor{orange}{\textbf{?}}} & \\ \hline
\cite{unbehauen-semantics-2016-sparqlmap-m} & \textcolor{checked}{\CheckmarkBold}/ & \textcolor{checked}{\CheckmarkBold}/\textcolor{checked}{\CheckmarkBold} & \textcolor{checked}{\CheckmarkBold} & \textcolor{checked}{\CheckmarkBold} & \textcolor{checked}{\CheckmarkBold} & \textcolor{checked}{\CheckmarkBold}/\textcolor{checked}{\CheckmarkBold} & {\Large \textcolor{orange}{\textbf{?}}} & /\textcolor{checked}{\CheckmarkBold} & {\Large \textcolor{orange}{\textbf{?}}} & \textcolor{checked}{\CheckmarkBold}/\textcolor{checked}{\CheckmarkBold} & \textcolor{checked}{\CheckmarkBold}/ & \\ \hline
\cite{mutharaju2013d} & {\Large \textcolor{orange}{\textbf{?}}} & \textcolor{red}{\XSolidBrush} & \textcolor{red}{\XSolidBrush} & {\Large \textcolor{orange}{\textbf{?}}} & \textcolor{red}{\XSolidBrush} & {\Large \textcolor{orange}{\textbf{?}}} & {\Large \textcolor{orange}{\textbf{?}}} &  & {\Large \textcolor{orange}{\textbf{?}}} & {\Large \textcolor{orange}{\textbf{?}}} & {\Large \textcolor{orange}{\textbf{?}}} & \\ \hline
\cite{botoeva2016obda} & {\Large \textcolor{orange}{\textbf{?}}} & \textcolor{checked}{\CheckmarkBold}/ & \textcolor{red}{\XSolidBrush} & {\Large \textcolor{orange}{\textbf{?}}} & \textcolor{red}{\XSolidBrush} & {\Large \textcolor{orange}{\textbf{?}}} & {\Large \textcolor{orange}{\textbf{?}}} & {\Large \textcolor{orange}{\textbf{?}}} & {\Large \textcolor{orange}{\textbf{?}}} & {\Large \textcolor{orange}{\textbf{?}}} & {\Large \textcolor{orange}{\textbf{?}}} & \\ \hline
\rowcolor{LightCyan}\multicolumn{13}{c}{\textit{SPARQL-to-XPath/XQuery}} \\ \hline
\cite{bikakis2015sparql2xquery,bikakis2009querying,bikakis2009semantic,bikakis2014supporting} & \textcolor{checked}{\CheckmarkBold}/\textcolor{checked}{\CheckmarkBold} & \textcolor{checked}{\CheckmarkBold}/\textcolor{checked}{\CheckmarkBold} & \textcolor{checked}{\CheckmarkBold} & \textcolor{checked}{\CheckmarkBold} & \textcolor{checked}{\CheckmarkBold} & \textcolor{checked}{\CheckmarkBold}/\textcolor{checked}{\CheckmarkBold}  & \textcolor{checked}{\CheckmarkBold} & {\Large \textcolor{orange}{\textbf{?}}} & {\Large \textcolor{orange}{\textbf{?}}} & \textcolor{checked}{\CheckmarkBold}/ & \textcolor{checked}{\CheckmarkBold}/\textcolor{checked}{\CheckmarkBold} & \texttt{DELETE}, \texttt{INSERT} \\ \hline
\cite{fischer2011translating} & \textcolor{checked}{\CheckmarkBold}/\textcolor{checked}{\CheckmarkBold} & \textcolor{checked}{\CheckmarkBold}/ & \textcolor{checked}{\CheckmarkBold} & \textcolor{checked}{\CheckmarkBold} & \textcolor{checked}{\CheckmarkBold} &  \textcolor{checked}{\CheckmarkBold}/\textcolor{checked}{\CheckmarkBold} & {\Large \textcolor{orange}{\textbf{?}}} & {\Large \textcolor{orange}{\textbf{?}}} & {\Large \textcolor{orange}{\textbf{?}}} & {\Large \textcolor{orange}{\textbf{?}}} & {\Large \textcolor{orange}{\textbf{?}}} &  \\ \hline
\cite{groppe2008embedding} & {\Large \textcolor{orange}{\textbf{?}}} & \textcolor{checked}{\CheckmarkBold}/\textcolor{checked}{\CheckmarkBold} & \textcolor{checked}{\CheckmarkBold} & \textcolor{checked}{\CheckmarkBold} & \textcolor{checked}{\CheckmarkBold} & {\Large \textcolor{orange}{\textbf{?}}} & {\Large \textcolor{orange}{\textbf{?}}} & {\Large \textcolor{orange}{\textbf{?}}} & {\Large \textcolor{orange}{\textbf{?}}} & {\Large \textcolor{orange}{\textbf{?}}} & {\Large \textcolor{orange}{\textbf{?}}} & \\ \hline
\end{tabular}
}
\caption{SPARQL features supported in SPARQL-to-X query translations. See Table~\ref{table:SQLCoverage} for \textcolor{checked}{\CheckmarkBold}~\textcolor{red}{\XSolidBrush}~{\Large \textcolor{orange}{\textbf{?}}}. \textit{Others} are features provided only by individual works.} \vspace{-15pt}
\label{table:SPARQLCoverage}
\end{table}

\subsection*{II. Translation Optimization} 
\subsubsection*{3. Optimization strategies} 
In this section, we use the terms previously introduced in \textit{Transformation type} (1); in order to avoid repetitions.

\subparagraph{XPath/XQuery-to-SQL:}
\cite{krishnamurthy2004efficient} suggests to eliminate joins by eliminating unnecessary prefix traversals, i.e. first traversals from the root. \cite{mani2006join} proposes a set of \textit{rewrite rules} meant to detect and eliminate unnecessarily redundant joins in the relational algebra of SQL queries resulted from the translation of XML queries. 
During query translation, \cite{fan2005query} suggests an algorithm  leveraging the structure of XML schema: pushing selections and projections into the LFP operator (Simple Least Fixpoint).
\textbf{PPFS+} \cite{georgiadis2007xpath} mainly seeks to leverage RDBMS storage of \textit{shredded} XML data. Based on an empirical evaluation, nested loop join was chosen to apply merge queries over the shredded XML. They try to improve query performance by generating \textit{pipelined} plans reducing time to "first results". 
To ensure XPath results follow the order of the original XML document and have as few duplicates as possible, redundant orders (ORDER BY) are eliminated, and ordering operations are pushed down the query plan tree. 
As a physical optimization, the article resorts to indexed file organization for the shredded relations. 
Even though \cite{min2008xtron} \textbf{XTRON} is schema-oblivious by nature, some schema/structural information is used to speed up query response. That is by encoding simple paths of XML elements into \textit{intervals} of real numbers using a specific algorithm (Reverse Arithmetic Encoder). The latter reduces the number of self-joins in the generated SQL queries. 

\subparagraph{SQL-to-XPath/XQuery:}
\textbf{ROX} \cite{halverson2004rox} suggests a cost-based optimization to generate optimal query plans, and physical indexes for quick node look-up; however, no details are given.

\subparagraph{SPARQL-to-SQL:}
The method in \cite{priyatna2014formalisation} optimizes certain SQL query cases that negatively impact (some) RDBMSs. In particular, \textit{sub-query elimination} and \textit{self-join elimination} query rewriting techniques are applied. The former removes non-correlated subqueries from the query by pushing down projections and selections, the latter removes self-joins occurring in the former queries. \cite{elliott2009complete} implements an optimization technique called "early project simplification", which skips variables that are not needed during query processing from the \texttt{SELECT} clause. 
In \textbf{SparqlMap} \cite{unbehauen2012accessing}, filter expressions are pushed to the graph patters, and nested SQL queries are flattened to minimize self-joins. 
In \textbf{FSparql2Sql} \cite{lu2008effective}, the translation method may generate an abnormal SQL query with a lot of CASE expressions and constants. The query is optimized by replacing complex expressions by simpler ones, e.g., by manipulating different logical orders, or removing useless ones. 
The translation approach in \textbf{Ultrawrap} \cite{sequeda2013ultrawrap} is expected to generate a view of a very large union of many SELECT-FROM-WHERE statements. To mitigate this, two strategies are applied: detection of unsatisfiable conditions, and self-join elimination. The former detects whether a query would yield empty results, even before executing it, due to the presence of contradictions e.g., \texttt{WHERE} predicate equals two opposite values; it also prunes unnecessary \texttt{UNION} sub-tree, e.g., by removing an empty argument from the \texttt{UNION}, in case two attributes of the same table are projected or filtered individually then joined. 
The generated SQL query in \cite{journals/ws/Rodriguez-MuroR15} may be sub-optimal due to the presence of e.g., joins of \texttt{UNION}-subqueries, redundant joins with respect to keys, unsatisfiable conditions. 
Using techniques from Logical Programming, \textit{Partial evaluation} is used to optimize Datalog rules dealing with \texttt{ans} and \texttt{triple} atoms, by iteratively filtering out options that would not generate valid answers; \textit{Goal Derivation} in Nested Atoms and Partial SDL-tree with \texttt{JOIN} and \texttt{LEFT JOIN} dealing with \textit{join} atoms.
Techniques from Semantic Query Optimizations are applied to detect unsatisfiable queries, e.g., joins when equating two different constants, simplification of trivially satisfiable conditions like $x=x$. 
The generated query in \cite{Chebotko06semanticspreserving} is optimized using \textit{simplifications}, e.g., removing redundant projections that do not contribute to a join or conditions in subqueries, removing \textit{True} values from some conditions, reducing join conditions based on logical evaluations, omitting left outer joins in case of SPARQL \texttt{UNION} when union'ed relations have identical schema, pushing down projection into \texttt{SELECT} subqueries, etc.

\subparagraph{SPARQL-to-Document:}
Query optimization in \textbf{D-SPARQ} \cite{mutharaju2013d} is based on a "divide and conquer"-like principle. It groups triple patterns into independent \textit{blocks} of triples, which can run more efficiently in parallel. For example, a star-shaped pattern groups are considered as indivisible blocks. Within one star pattern group, for each predicate triple patterns are ordered by number of triples involving that predicate. This boosts query processing by reducing the selectivity of the individual patter groups. 
In the relational-based OBDA of \cite{botoeva2016obda}, the intermediate relational query is simplified by applying structural optimization, e.g., replacing join of unions by union of joins, and semantic optimization, e.g., redundant self-join elimination.
In \cite{michel2016generic}, the generated MongoDB query is optimized by pushing filters to the level of triple patters, and by self-join elimination through merging atomic queries that share the same \texttt{FROM} part, and by self-union elimination through merging \texttt{UNION}s of atomic queries that share the same \texttt{FROM} part. 

\subparagraph{Cypher-to-SQL:}
\textbf{Cyp2sql} \cite{cyp2sql} stores graph data following a specific tables scheme, which is designed to optimize specific queries. For example, \textit{Label} table is created to overcome the problem of prevalent NULL values in the Nodes table. Query translator decides, on \textit{query-time}, which relationship to use to obtain node information. Relationship data is stored in the \textit{Edges} table (storing all relationships) as well as in their separate tables (duplicate). Further optimization is gained from using a couple of metafiles populated during schema conversion, e.g., a nodes property list per label type used to narrow down the search for nodes.

\subparagraph{SPARQL-to-XPath/XQuery:}
In \cite{groppe2008embedding}, a logical optimization is applied to the operator tree in order to generate a reorganized equivalent tree with faster translation time (no more details given). Next, a physical optimization aims to find the algorithm that implements the operator with the best estimated performance. 

\subparagraph{Gremlin-to-SQL}
SQLGraph~\cite{sun2015sqlgraph} proposes a translation optimization whereby a sequence of the non selective pipe g.V (retrieve all vertices in g) or g.E (retrieve all edges in g) are replaced by a sequence of attribute-based filter pipes (filter pipes that select graph elements based on specific values). For example, the non selective first pipe g.V is explicitly merged with the more selective filter \texttt{filter{it.tag == 'w'}} in the translation. For the query evaluation, optimization strategies of the RDBMS are leveraged.


\subsubsection*{4. Translation relationship}
This information is not always explicitly stated, and we cannot make assumptions based on the architectures or the algorithms, so we only report when there is a clear statement about the type of relationship. Information is collected in Table~\ref{table:relationship}.
\begin{table}[t]
\centering
\begin{tabular}{|l|c|c|}
\hline
\rowcolor{blue!35}\textbf{Work} & \textbf{One-to-one} & \textbf{One-to-many} \\ \hline
\rowcolor{LightCyan}\multicolumn{3}{c}{\textit{SQL-to-XPath/XQuery:}} \\ \hline
\cite{halverson2004rox} ROX &  & \textcolor{checked}{\CheckmarkBold} \\ \hline
\rowcolor{LightCyan}\multicolumn{3}{c}{\textit{SPARQL-to-SQL:}} \\ \hline
\cite{kiminki2010sparql} Type-ARQuE & \textcolor{checked}{\CheckmarkBold} & \\ \hline
\cite{lu2008effective} FSparql2Sql & \textcolor{checked}{\CheckmarkBold} &  \\ \hline
\rowcolor{LightCyan}\multicolumn{3}{c}{\textit{SQL-to-SPARQL:}} \\ \hline
\cite{ramanujam2009r2d,ramanujam2009r2dextra} R2D \textit{SQL-to-SPARQL:}
 & \textcolor{checked}{\CheckmarkBold} &  \\ \hline
\rowcolor{LightCyan}\multicolumn{3}{c}{\textit{SQL-to-Document-based:}} \\ \hline
\cite{querymongo} QueryMongo & \textcolor{checked}{\CheckmarkBold} & \\ \hline
\rowcolor{LightCyan}\multicolumn{3}{c}{\textit{Gremlin-to-SQL:}} \\ \hline
\cite{sun2015sqlgraph} SQLGraph & \textcolor{checked}{\CheckmarkBold} &  \\ \hline
\end{tabular}
\caption{Query Translation relationship.} \vspace{-20pt}
\label{table:relationship}
\end{table}

    
\subsection*{III. Community Factors}
For a better readability and structuring, we collect the information in Table~\ref{table:communityFactors}. The last column rates the community effect using stars (\textcolor{NiceGold}{\FiveStar}), which are to be interpreted as follows. \textcolor{NiceGold}{\FiveStar}: `Implemented', 
\textcolor{NiceGold}{\FiveStar}\textcolor{NiceGold}{\FiveStar}: `Implemented and Evaluated' or `Implemented and Available (for download)', `\textcolor{NiceGold}{\FiveStar}\textcolor{NiceGold}{\FiveStar}\textcolor{NiceGold}{\FiveStar}: `Implemented, Evaluated and Available (for download)'.
\begin{table}[p]
\resizebox{\textwidth}{!}{
\Large
\begin{tabular}{|p{5cm}|cccc|c|c|}
\hline
\rowcolor{blue!35}\textbf{Paper/tool} & $\mathbf{Y_{FR}}$ & $\mathbf{Y_{LR}}$ & $\mathbf{n_R}$ & $\mathbf{n_C}$ & \textbf{Implementation Reference} & \textbf{Community} \\ \hline
\rowcolor{LightCyan}\multicolumn{7}{c}{\textit{XPath/XQuery-to-SQL}} \\ \hline
\cite{krishnamurthy2004efficient} & & & & 57 & & \textcolor{NiceGold}{\FiveStar} \\ \hline
\cite{fan2005query} & 2005 & & & 37 & & \textcolor{NiceGold}{\FiveStar}\textcolor{NiceGold}{\FiveStar} \\ \hline
\cite{mani2006join} & 2006 & & 1 & & & \textcolor{NiceGold}{\FiveStar}\textcolor{NiceGold}{\FiveStar} \\ \hline
\cite{georgiadis2007xpath} PPFS+ & & & & 40 & & \textcolor{NiceGold}{\FiveStar}\textcolor{NiceGold}{\FiveStar} \\ \hline
\cite{hu2008adaptive} & & & & 5 & & \textcolor{NiceGold}{\FiveStar}\textcolor{NiceGold}{\FiveStar} \\ \hline
\cite{min2008xtron} XTRON & & & & 23 & & \textcolor{NiceGold}{\FiveStar}\textcolor{NiceGold}{\FiveStar} \\ \hline
\rowcolor{LightCyan}\multicolumn{7}{c}{\textit{SQL-to-XPath/XQuery}} \\ \hline
\cite{vidhya2009query, vidhya2010insert} & & & & 1, 5 & & \\ \hline
\cite{jigyasu2006sql} AquaLogic & 2006 & 2008 & & 22 & \textit{Acquired by Oracle and merged in its products} & \textcolor{NiceGold}{\FiveStar}\textcolor{NiceGold}{\FiveStar} \\ \hline
\cite{halverson2004rox} & & & & 65 & & \textcolor{NiceGold}{\FiveStar}\textcolor{NiceGold}{\FiveStar} \\ \hline
\rowcolor{LightCyan}\multicolumn{7}{c}{\textit{SPARQL-to-SQL}} \\ \hline
\cite{stadlerconnecting} Sparqlify  & 2013 & 2018 & 30 & 2 & \url{https://github.com/SmartDataAnalytics/Sparqlify} & \textcolor{NiceGold}{\FiveStar}\textcolor{NiceGold}{\FiveStar}\textcolor{NiceGold}{\FiveStar} \\ \hline
\cite{kiminki2010sparql} Type-ARQuE & & 2010 & & 6 & \url{http://www.cs.hut.fi/~skiminki/type-arque/index.html} & \textcolor{NiceGold}{\FiveStar}\textcolor{NiceGold}{\FiveStar} \\ \hline
\cite{priyatna2014formalisation} Morph translator & 2014 & 2018 & 37 & 74 & \textit{Part of Morph-RDB:} \url{https://github.com/oeg-upm/morph-rdb} & \textcolor{NiceGold}{\FiveStar}\textcolor{NiceGold}{\FiveStar}\textcolor{NiceGold}{\FiveStar} \\ \hline
\cite{Chebotko06semanticspreserving} & & & & 151 & & \textcolor{NiceGold}{\FiveStar}\textcolor{NiceGold}{\FiveStar} \\ \hline
\cite{lu2008effective} & & & & 28 & & \textcolor{NiceGold}{\FiveStar}\textcolor{NiceGold}{\FiveStar} \\ \hline
\cite{elliott2009complete} & & & & 78 & & \textcolor{NiceGold}{\FiveStar}\textcolor{NiceGold}{\FiveStar}\textcolor{NiceGold}{\FiveStar} \\ \hline
\cite{unbehauen2012accessing} SPARQLMap & & & & 22 & & \textcolor{NiceGold}{\FiveStar}\textcolor{NiceGold}{\FiveStar}\textcolor{NiceGold}{\FiveStar} \\ \hline
\cite{sequeda2013ultrawrap} Ultrawrap& & & & 99 & https://capsenta.com/ultrawrap & \textcolor{NiceGold}{\FiveStar}\textcolor{NiceGold}{\FiveStar} \\ \hline
\cite{journals/ws/Rodriguez-MuroR15} & & & & 52 & \textit{Part of Ontop: https://github.com/ontop/ontop} & \textcolor{NiceGold}{\FiveStar}\textcolor{NiceGold}{\FiveStar} \\ \hline
\rowcolor{LightCyan}\multicolumn{7}{c}{\textit{SQL-to-SPARQL}} \\ \hline
\cite{ramanujam2009r2d,ramanujam2009r2dextra} R2D & & & & 19, 15 & & \textcolor{NiceGold}{\FiveStar}\textcolor{NiceGold}{\FiveStar} \\ \hline
\cite{rachapalli2011retro} & & & & 14 & & \\ \hline
\rowcolor{LightCyan}\multicolumn{7}{c}{SQL-to-Document-based} \\ \hline
\cite{querymongo} Query Mongo & & & & & & \\ \hline
\cite{mongoDB-translator-teiid} MongoDB Translator & & & & & & \textcolor{NiceGold}{\FiveStar} \\ \hline
\cite{unityjdbc} UnityJDBC & & & & & & \textcolor{NiceGold}{\FiveStar}\\ \hline
\rowcolor{LightCyan}\multicolumn{7}{c}{SPARQL-to-Document-based} \\ \hline
\cite{mutharaju2013d} D-SPARQ & & & & 11 & & \textcolor{NiceGold}{\FiveStar}\textcolor{NiceGold}{\FiveStar}\\ \hline
\cite{unbehauen-semantics-2016-sparqlmap-m} SparqlMap-M & 2015 & 2017 & 12 & 2 & \url{https://github.com/tomatophantastico/sparqlmap} & \textcolor{NiceGold}{\FiveStar}\textcolor{NiceGold}{\FiveStar}\textcolor{NiceGold}{\FiveStar} \\ \hline
\cite{botoeva2016obda} & & & & 19 & \textit{Extends Ontop but no reference found} & \textcolor{NiceGold}{\FiveStar} \\ \hline
\cite{conf/ontobras/AraujoABW17} OntoMongo & & 2017 & & 1 & \url{https://github.com/thdaraujo/onto-mongo} & \textcolor{NiceGold}{\FiveStar}\textcolor{NiceGold}{\FiveStar} \\ \hline
\cite{michel2016generic} & 2014 & 2015 & 6 & 5 & \url{https://github.com/frmichel/morph-xr2rml/tree/query\_rewrite} & \textcolor{NiceGold}{\FiveStar}\textcolor{NiceGold}{\FiveStar} \\ \hline
\rowcolor{LightCyan}\multicolumn{7}{c}{\textit{\textit{Cypher-to-SQL}}} \\ \hline
\cite{steer2017cytosm} Cytosm & & 2017 & 1 & 2 & \url{https://github.com/cytosm/cytosm} & \textcolor{NiceGold}{\FiveStar}\textcolor{NiceGold}{\FiveStar}\textcolor{NiceGold}{\FiveStar} \\ \hline
\cite{cyp2sql} Cyp2sql & 2017 & 2017 & & 1 & \url{https://github.com/DTG-FRESCO/cyp2sql} & \textcolor{NiceGold}{\FiveStar}\textcolor{NiceGold}{\FiveStar} \\  \hline
\rowcolor{LightCyan}\multicolumn{7}{c}{\textit{Gremlin-to-SQL}} \\ \hline
\cite{sun2015sqlgraph} SQLGraph & 2015 & & & 44 & & \textcolor{NiceGold}{\FiveStar}\textcolor{NiceGold}{\FiveStar} \\ \hline
\rowcolor{LightCyan}\multicolumn{7}{c}{\textit{SQL-to-Gremlin}} \\ \hline
\cite{sqlgremlin} SQL-Gremlin & 2015 & 2016 & 1 & & \url{https://github.com/twilmes/sql-gremlin} & \textcolor{NiceGold}{\FiveStar} \\ \hline
\rowcolor{LightCyan}\multicolumn{7}{c}{\textit{SPARQL-to-XPath/XQuery}} \\ \hline
\cite{bikakis2015sparql2xquery,bikakis2009querying,bikakis2009semantic} SPARQL2XQuery & & & & 29, 11, 21 & \url{http://www.dblab.ntua.gr/~bikakis/SPARQL2XQuery.html} & \textcolor{NiceGold}{\FiveStar}\textcolor{NiceGold}{\FiveStar}\textcolor{NiceGold}{\FiveStar} \\ \hline

\cite{groppe2008embedding} & & & & 45 & & \textcolor{NiceGold}{\FiveStar}\textcolor{NiceGold}{\FiveStar}\\ \hline
\cite{fischer2011translating} XQL2Xquery & & & & 6 & & \textcolor{NiceGold}{\FiveStar}\textcolor{NiceGold}{\FiveStar} \\ \hline
\rowcolor{LightCyan}\multicolumn{7}{c}{\textit{XPath/XQuery-to-SPARQL}} \\ \hline
\cite{droop2007translating} & & & & 21 & & \textcolor{NiceGold}{\FiveStar}\textcolor{NiceGold}{\FiveStar} \\ \hline
\rowcolor{LightCyan}\multicolumn{7}{c}{\textit{SPARQL-to-Gremlin}} \\ \hline
\cite{thakkar2018stitch,thakkar2018two} Gremlinator & 2018 & & & 6 & \url{https://github.com/apache/tinkerpop/tree/master/sparql-gremlin} & \textcolor{NiceGold}{\FiveStar}\textcolor{NiceGold}{\FiveStar}\textcolor{NiceGold}{\FiveStar} \\ \hline
\end{tabular}
}
\caption{Community Factors. $\mathbf{Y_{FR}}$ year of first release, $\mathbf{Y_{LR}}$ year of last release, $\mathbf{n_R}$ number of releases, $\mathbf{n_C}$ number of citations (from Google Scholar). If $\mathbf{n_R}=1$ it is the first release and last release is last update.} 
\label{table:communityFactors}
\end{table}

\section*{Discussions and Conclusion}
\label{sec:discussion}
\paragraph{Weakly addressed paths.}
Although one would presume that \textit{SQL-to-Document-based} translation is a well-supported path given the popularity of SQL and document databases, there is still a modest literature in this regard. Most of the efforts provide marginal contributions in addition to the more general SQL-to-NoSQL translation. Furthermore, the translation of this path in all cases is far from being complete, and does not follow the systematic methodology observed by other efforts in this study. Some of these works are \cite{dos2013providing,schreiner2015sqltokeynosql,lawrence2014integration}. 
Similarly, despite the popularity of SQL and Gremlin, the \textit{Gremlin-to-SQL} translation has also attracted little attention. That may be due to the large difference in the semantics of the Gremlin graph traversal model and SQL's relational model. 
In general, the work on translating between SQL and MongoDB and Gremlin languages is still in an relatively early stage, partially because of the lack of a strong formal foundation of the semantics and complexity of MongoDB's document language as well as Gremlin.
On the other hand, the path \textit{XPath/XQuery-to-SPARQL} has much fewer works than its reverse. This is possibly because SPARQL is more frequently used for solving integration problems as part of the OBDA framework, which involves translating various queries into SPARQL.

\begin{table}[p]
\centering
\begin{minipage}[t]{\linewidth}
\color{gray}
\rule{\linewidth}{1pt}
\ytl{1974}{\citeauthor{chamberlin1974sequel}}{\textit{SQL introduced}}
\ytl{2002}{\citeauthor{boag2002xquery}}{\textit{XQuery introduced}}
\ytl{2003}{\citeauthor{berglund2003xml}}{\textit{XPath introduced}}
\ytl{2004}{\citeauthor{halverson2004rox} \textbf{ROX} \\ \cite{krishnamurthy2004efficient,krishnamurthy2004recursive}}{SQL-to-XPath/XQuery \\ XPath/XQuery-to-SQL}
\ytl{2005}{\citeauthor{fan2005query}}{XPath-to-SQL}
\ytl{2006}{\citeauthor{mani2006join}}{XQuery-to-SQL} 
\ytl{2007}{\citeauthor{droop2007translating} \\ \citeauthor{georgiadis2007xpath}}{XPath/XQuery-to-SPARQL \\ XPath/XQuery-to-SQL}
\ytl{2008}{\citeauthor{prudhommeaux2008sparql} \\ \citeauthor{hu2008adaptive} \\ \citeauthor{lu2008effective} \\ \citeauthor{min2008xtron} \textbf{XTRON}}{\textit{SPARQL introduced} \\ XML-to-SQL \\ SPARQL-to-SQL \\ XQuery-to-SQL}
\ytl{2009}{\citeauthor{fan2009query} \\ \citeauthor{vidhya2009query} \\ \citeauthor{elliott2009complete} \\ \citeauthor{bikakis2009querying,bikakis2009semantic} \\ \citeauthor{ramanujam2009r2d}}{XPath-to-SQL \\ SQL-to-XPath \\ SPARQL-to-SQL \\ SPARQL-to-XQuery \\ SQL-to-SPARQL}
\ytl{2010}{\citeauthor{vidhya2010insert} \\ \citeauthor{kiminki2010sparql} \textbf{Type-ARQuE}}{SQL-to-XQuery \\ SPARQL-to-SQL}
\ytl{2011}{\citeauthor{das2011r2rml} \textbf{R2RML} \\ \citeauthor{atay2011schema} \\ \citeauthor{fischer2011translating} \\ \citeauthor{rachapalli2011retro} \textbf{RETRO}}{SQL-to-SPARQL \\ XML-to-SQL \\ SQL- and SPARQL-to-XQuery \\ SQL-to-SPARQL}
\ytl{2012}{\citeauthor{rodriguez2012quest} \textbf{Quest} \\ \citeauthor{unbehauen2012accessing} \textbf{SPARQLMap}}{SPARQL-to-SQL \\ SPARQL-to-SQL}
\ytl{2013}{\citeauthor{dos2013providing} \\ \citeauthor{sequeda2013ultrawrap} \textbf{Ultrawrap}}{SQL-to-Document based \\ SPARQL-to-SQL}
\ytl{2014}{\citeauthor{bikakis2014supporting} \\ \citeauthor{priyatna2014formalisation} \textbf{Morph} \\ \citeauthor{lawrence2014integration}}{SPARQL-to-XQuery \\ SPARQL-to-SQL \\ SQL-to-Document-based}
\ytl{2015}{\citeauthor{sun2015sqlgraph} \textbf{SQLGraph} \\ \citeauthor{bikakis2015sparql2xquery}}{Gremlin-to-SQL \\ SPARQL-to-XQuery}
\ytl{2016}{\citeauthor{unbehauen-semantics-2016-sparqlmap-m} \textbf{SparqlMap-M}}{SQL-to-Document-based}
\ytl{2017}{\citeauthor{steer2017cytosm} \textbf{Cytosm}}{Cypher-to-SQL} \ytl{2018}{\citeauthor{thakkar2018stitch,thakkar2018two} \textbf{Gremlinator}}{SPARQL-to-Gremlin}
\end{minipage}%
\caption{Timeline recording publication years of the considered query languages and methods.}
\label{table:history}
\end{table}

\paragraph{Missing paths.}
We have not found any articles or software/tools for the following paths \textit{SQL-to-Cypher}, \textit{Gremlin-to-SPARQL}, \textit{XPath/XQuery-to-Cypher} and vice versa, \textit{XPath/XQuery-to-Gremlin} and vice versa, \textit{Cypher-to-Document-based} and vice versa.
We see opportunities in tackling those translation paths with rationals similar to those of the previously tackled translation paths. For example, although SPARQL and Gremlin fundamentally differ in their approaches to query graph data, one based on graph pattern matching one on graph traversals, they are both graph query languages. A transition from one to the other not only allows the interoperability between systems supporting those languages, but also makes data from one world available to the other without requiring to learn the other respective query language~\cite{DBLP:conf/amw/AnglesTT19}. Similarly, XML languages have a rooted notion of traversals, a conversion to and from Gremlin is natural. In fact, according to~\cite{lindaaker2018graphhistory}, the early prototype of Gremlin used XPath for querying graph data.

\paragraph{Gaps and Lessons Learned.}
The survey has also allowed us to identify gaps and learn lessons, which we summarize in the following points:
\begin{itemize}[nosep]
    \item We noticed that the optimizations that are applied during the query translation process have more potential to improve the overall translation performance than the optimization applied on the generated query. This is because at query translation-time, optimizations from the system of the original query, e.g., statistics, can be leveraged to impact the resulted target query. This opportunity is not present once the query in the target language has been generated.
    \item Looking at the language scope coverage, there seems to still be a lack in covering the more sophisticated operations of query languages, e.g., more join types and temporal functions in SQL; blank nodes, grouping and binding in SPARQL. Such functions are motivated by and are at the core of many modern analytical and real-time applications. Indeed, some of those features are newly-introduced and some of the needs are only recently exposed, in which case we make the call to both update the existing works and build new solutions to embrace the new features and address the new needs.
    \item Certain works present a well-founded and defined query translation frameworks, from the query translation process to the various optimization strategies. However, the example queries effectively worked on are simple and would hardly represent real-world queries. Use-case-driven translation methods would be more helpful to reveal the useful query patterns and fragments, and to evaluate the translation methods and optimizations on real-world data.
    \item There is a wide variety in the evaluation frameworks used by each of the query translation methods. Following a unique standardized benchmark specialized in evaluating and assessing query translation aspects is paramount. Such a dedicated benchmark unfortunately does not exist at the time of writing.
\end{itemize}

\paragraph{Candidates for a 'universal' query language.} After discovering and exploring the various query translation methods, it appears that SQL and SPARQL are the most suitable languages to act as a 'universal' language for realizing the heterogeneous data integration. They both have the most number of translations to other languages (see outgoing edges in Figure~\ref{fig:translationPaths}). SQL is the oldest query language with ever-continued development cycles and adoption. SPARQL is the stable query language of the so-called ontology-based data integration and access, which specializes specifically in integrating data coming from heterogeneous sources.

\paragraph{Query Translation History.} We project the surveyed works into a vertical timeline shown in Table~\ref{table:history}. The visualization allows us to draw some remarks. SPARQL was very quickly recognized by the community, as works translating to and from SPARQL started to emerge the same year it was suggested. We cannot make a similar judgment about the adoption of SQL, XPath and XQuery as they were introduced earlier than the timeframe we consider in this study, 2003-2019. Works on translating to and from SPARQL have continued to attract research efforts to date. Works translating to and from SQL is present in all the years of the timeline, except 2013. With less regularity, works translating to and from XML languages have also been continually published. Despite their latest updates in 2017, we have not found any works (at least complying with our criteria) published since 2015.

In this article, we have surveyed more than forty articles and tools around query translation between seven popular query languages. Although organizing the information was a complicated and sensitive task, the study allowed us to extract eight common criteria according to which we categorized the surveyed works. It also allowed us to discover which translation paths are not sufficiently addressed and which ones are not addressed yet, as well as to observe gaps and learn lessons for future research on the topic. 
We hope that reporting this knowledge opens new doors for research and development on the topic of query translation, and serves users of applications like polyglot persistence and data lakes to exploit more data value by tackling the data variety issue.


\bibliography{main}

\begin{thebibliography}{92}
\providecommand{\natexlab}[1]{#1}
\providecommand{\url}[1]{\texttt{#1}}
\expandafter\ifx\csname urlstyle\endcsname\relax
  \providecommand{\doi}[1]{doi: #1}\else
  \providecommand{\doi}{doi: \begingroup \urlstyle{rm}\Url}\fi

\bibitem[Angles et~al.(2017)Angles, Arenas, Barcel{\'o}, Hogan, Reutter, and
  Vrgo{\v{c}}]{angles2017foundations}
Renzo Angles, Marcelo Arenas, Pablo Barcel{\'o}, Aidan Hogan, Juan Reutter, and
  Domagoj Vrgo{\v{c}}.
\newblock Foundations of modern query languages for graph databases.
\newblock \emph{ACM Computing Surveys (CSUR)}, 50\penalty0 (5):\penalty0 68,
  2017.

\bibitem[Angles et~al.(2019)Angles, Thakkar, and
  Tomaszuk]{DBLP:conf/amw/AnglesTT19}
Renzo Angles, Harsh Thakkar, and Dominik Tomaszuk.
\newblock {RDF} and {P}roperty {G}raphs {I}nteroperability: {S}tatus and
  {I}ssues.
\newblock In \emph{Proceedings of the 13th Alberto Mendelzon International
  Workshop on Foundations of Data Management, Asunci{\'{o}}n, Paraguay, June
  3-7, 2019.}, 2019.
\newblock URL \url{http://ceur-ws.org/Vol-2369/paper01.pdf}.

\bibitem[Araujo et~al.(2017)Araujo, Agena, Braghetto, and
  Wassermann]{conf/ontobras/AraujoABW17}
Thiago Henrique~Dias Araujo, Barbara~Tieko Agena, Kelly~Rosa Braghetto, and
  Renata Wassermann.
\newblock Ontomongo- ontology-based data access for nosql.
\newblock In Mara Abel, Sandro~Rama Fiorini, and Christiano Pessanha, editors,
  \emph{ONTOBRAS}, volume 1908 of \emph{CEUR Workshop Proceedings}, pages
  55--66. CEUR-WS.org, 2017.
\newblock URL \url{http://dblp.uni-trier.de/db/conf/ontobras/
  ontobras2017.html#AraujoABW17}.

\bibitem[Atay and Chebotko(2011)]{atay2011schema}
Mustafa Atay and Artem Chebotko.
\newblock Schema-based xml-to-sql query translation using interval encoding.
\newblock In \emph{Information Technology: New Generations (ITNG), 2011 Eighth
  International Conference on}, pages 84--89. IEEE, 2011.

\bibitem[Berglund et~al.(2003)Berglund, Boag, Chamberlin, Fern{\'a}ndez, Kay,
  Robie, and Sim{\'e}on]{berglund2003xml}
Anders Berglund, Scott Boag, Don Chamberlin, Mary~F Fern{\'a}ndez, Michael Kay,
  Jonathan Robie, and J{\'e}r{\^o}me Sim{\'e}on.
\newblock Xml path language (xpath).
\newblock \emph{World Wide Web Consortium (W3C)}, 2003.

\bibitem[Bikakis et~al.(2009{\natexlab{a}})Bikakis, Gioldasis, Tsinaraki, and
  Christodoulakis]{bikakis2009querying}
Nikos Bikakis, Nektarios Gioldasis, Chrisa Tsinaraki, and Stavros
  Christodoulakis.
\newblock Querying xml data with sparql.
\newblock In \emph{International Conference on Database and Expert Systems
  Applications}, pages 372--381. Springer, 2009{\natexlab{a}}.

\bibitem[Bikakis et~al.(2009{\natexlab{b}})Bikakis, Gioldasis, Tsinaraki, and
  Christodoulakis]{bikakis2009semantic}
Nikos Bikakis, Nektarios Gioldasis, Chrisa Tsinaraki, and Stavros
  Christodoulakis.
\newblock Semantic based access over xml data.
\newblock In \emph{World Summit on Knowledge Society}, pages 259--267.
  Springer, 2009{\natexlab{b}}.

\bibitem[Bikakis et~al.(2014)Bikakis, Tsinaraki, Stavrakantonakis, and
  Christodoulakis]{bikakis2014supporting}
Nikos Bikakis, Chrisa Tsinaraki, Ioannis Stavrakantonakis, and Stavros
  Christodoulakis.
\newblock Supporting sparql update queries in rdf-xml integration.
\newblock \emph{arXiv preprint arXiv:1408.2800}, 2014.

\bibitem[Bikakis et~al.(2015)Bikakis, Tsinaraki, Stavrakantonakis, Gioldasis,
  and Christodoulakis]{bikakis2015sparql2xquery}
Nikos Bikakis, Chrisa Tsinaraki, Ioannis Stavrakantonakis, Nektarios Gioldasis,
  and Stavros Christodoulakis.
\newblock The sparql2xquery interoperability framework.
\newblock \emph{World Wide Web}, 18\penalty0 (2):\penalty0 403--490, 2015.

\bibitem[Boag et~al.(2002)Boag, Chamberlin, Fern{\'a}ndez, Florescu, Robie,
  Sim{\'e}on, and Stefanescu]{boag2002xquery}
Scott Boag, Don Chamberlin, Mary~F Fern{\'a}ndez, Daniela Florescu, Jonathan
  Robie, J{\'e}r{\^o}me Sim{\'e}on, and Mugur Stefanescu.
\newblock Xquery 1.0: An xml query language.
\newblock 2002.

\bibitem[Botoeva et~al.(2016)Botoeva, Calvanese, Cogrel, Rezk, and
  Xiao]{botoeva2016obda}
Elena Botoeva, Diego Calvanese, Benjamin Cogrel, Martin Rezk, and Guohui Xiao.
\newblock Obda beyond relational dbs: A study for mongodb.
\newblock CEUR Workshop Proceedings, 2016.

\bibitem[Bray et~al.(1997)Bray, Paoli, Sperberg-McQueen, Maler, and
  Yergeau]{bray1997extensible}
Tim Bray, Jean Paoli, C~Michael Sperberg-McQueen, Eve Maler, and Fran{\c{c}}ois
  Yergeau.
\newblock Extensible markup language (xml).
\newblock \emph{World Wide Web Journal}, 2\penalty0 (4):\penalty0 27--66, 1997.

\bibitem[Carata(2017)]{cyp2sql}
Lucian Carata.
\newblock Cyp2sql: Cypher to sql translation, 2017.
\newblock URL \url{https://github.com/DTG-FRESCO/cyp2sql}.
\newblock Accessed: 06-08-2019.

\bibitem[Chamberlin and Boyce(1974)]{chamberlin1974sequel}
Donald~D Chamberlin and Raymond~F Boyce.
\newblock Sequel: A structured english query language.
\newblock In \emph{Proceedings of the 1974 ACM SIGFIDET (now SIGMOD) workshop
  on Data description, access and control}, pages 249--264. ACM, 1974.

\bibitem[Chebotko et~al.(2006)Chebotko, Lu, Jamil, and
  Fotouhi]{Chebotko06semanticspreserving}
Artem Chebotko, Shiyong Lu, Hasan~M. Jamil, and Farshad Fotouhi.
\newblock Semantics preserving sparql-to-sql query translation for optional
  graph patterns.
\newblock Technical report, 2006.

\bibitem[Clark et~al.(1999)Clark, DeRose, et~al.]{clark1999xml}
James Clark, Steve DeRose, et~al.
\newblock Xml path language (xpath) version 1.0, 1999.

\bibitem[Codd(1970)]{codd1970relational}
Edgar~F Codd.
\newblock A relational model of data for large shared data banks.
\newblock \emph{Communications of the ACM}, 13\penalty0 (6):\penalty0 377--387,
  1970.

\bibitem[Das(2011)]{das2011r2rml}
Souripriya Das.
\newblock R2rml: Rdb to rdf mapping language.
\newblock \emph{http://www. w3. org/TR/r2rml/}, 2011.

\bibitem[Dixon(2010)]{2010dixon}
James Dixon.
\newblock {Pentaho, Hadoop, and Data Lakes}, 2010.
\newblock URL
  \url{https://jamesdixon.wordpress.com/2010/10/14/pentaho-hadoop-and-data-lakes}.
\newblock Online; accessed 27-January-2019.

\bibitem[dos Santos~Ferreira et~al.(2013)dos Santos~Ferreira, Calil, and dos
  Santos~Mello]{dos2013providing}
Gabriel dos Santos~Ferreira, Andre Calil, and Ronaldo dos Santos~Mello.
\newblock On providing ddl support for a relational layer over a document nosql
  database.
\newblock In \emph{Proceedings of International Conference on Information
  Integration and Web-based Applications \& Services}, page 125. ACM, 2013.

\bibitem[Droop et~al.(2007)Droop, Flarer, Groppe, Groppe, Linnemann, Pinggera,
  Santner, Schier, Sch{\"o}pf, Staffler, et~al.]{droop2007translating}
Matthias Droop, Markus Flarer, Jinghua Groppe, Sven Groppe, Volker Linnemann,
  Jakob Pinggera, Florian Santner, Michael Schier, Felix Sch{\"o}pf, Hannes
  Staffler, et~al.
\newblock Translating xpath queries into sparql queries.
\newblock In \emph{OTM Confederated International Conferences" On the Move to
  Meaningful Internet Systems"}, pages 9--10. Springer, 2007.

\bibitem[Dyck et~al.(2017)Dyck, Robie, Snelson, and Chamberlin]{dyck2017xml}
Michael Dyck, J~Robie, J~Snelson, and D~Chamberlin.
\newblock Xml path language (xpath) 3.1, 2017.
\newblock URL \url{https://www.w3.org/TR/xpath-31/}.
\newblock Accessed: 06-08-2019.

\bibitem[Elliott et~al.(2009)Elliott, Cheng, Thomas-Ogbuji, and
  Ozsoyoglu]{elliott2009complete}
Brendan Elliott, En~Cheng, Chimezie Thomas-Ogbuji, and Z~Meral Ozsoyoglu.
\newblock A complete translation from sparql into efficient sql.
\newblock In \emph{Proceedings of the 2009 International Database Engineering
  \& Applications Symposium}, pages 31--42. ACM, 2009.

\bibitem[Fan et~al.(2005)Fan, Yu, Lu, Lu, and Rastogi]{fan2005query}
Wenfei Fan, Jeffrey~Xu Yu, Hongjun Lu, Jianhua Lu, and Rajeev Rastogi.
\newblock Query translation from xpath to sql in the presence of recursive
  dtds.
\newblock In \emph{Proceedings of the 31st international conference on Very
  large data bases}, pages 337--348. VLDB Endowment, 2005.

\bibitem[Fan et~al.(2009)Fan, Yu, Li, Ding, and Qin]{fan2009query}
Wenfei Fan, Jeffrey~Xu Yu, Jianzhong Li, Bolin Ding, and Lu~Qin.
\newblock Query translation from xpath to sql in the presence of recursive
  dtds.
\newblock \emph{The VLDB Journal}, 18\penalty0 (4):\penalty0 857--883, 2009.

\bibitem[Fischer et~al.(2011)Fischer, Florescu, Kaufmann, and
  Kossmann]{fischer2011translating}
Peter~M Fischer, Dana Florescu, Martin Kaufmann, and Donald Kossmann.
\newblock Translating sparql and sql to xquery.
\newblock \emph{XML Prague}, pages 81--98, 2011.

\bibitem[Francis et~al.(2018)Francis, Green, Guagliardo, Libkin, Lindaaker,
  Marsault, Plantikow, Rydberg, Selmer, and Taylor]{francis2018cypher}
Nadime Francis, Alastair Green, Paolo Guagliardo, Leonid Libkin, Tobias
  Lindaaker, Victor Marsault, Stefan Plantikow, Mats Rydberg, Petra Selmer, and
  Andr{\'e}s Taylor.
\newblock Cypher: An evolving query language for property graphs.
\newblock In \emph{Proceedings of the 2018 International Conference on
  Management of Data}, pages 1433--1445. ACM, 2018.

\bibitem[Gearon et~al.(2013)Gearon, Passant, and Polleres]{gearon2013sparql}
Paul Gearon, Alexandre Passant, and Axel Polleres.
\newblock Sparql 1.1 update.
\newblock \emph{W3C recommendation}, 21, 2013.

\bibitem[Georgiadis and Vassalos(2007)]{georgiadis2007xpath}
Haris Georgiadis and Vasilis Vassalos.
\newblock Xpath on steroids: exploiting relational engines for xpath
  performance.
\newblock In \emph{Proceedings of the 2007 ACM SIGMOD international conference
  on Management of data}, pages 317--328. ACM, 2007.

\bibitem[Green et~al.(2018)Green, Junghanns, Kiessling, Lindaaker, Plantikow,
  and Selmer]{green2018opencypher}
Alastair Green, Martin Junghanns, Max Kiessling, Tobias Lindaaker, Stefan
  Plantikow, and Petra Selmer.
\newblock opencypher: New directions in property graph querying.
\newblock 2018.

\bibitem[Groppe et~al.(2008)Groppe, Groppe, Linnemann, Kukulenz, Hoeller, and
  Reinke]{groppe2008embedding}
Sven Groppe, Jinghua Groppe, Volker Linnemann, Dirk Kukulenz, Nils Hoeller, and
  Christoph Reinke.
\newblock Embedding sparql into xquery/xslt.
\newblock In \emph{Proceedings of the 2008 ACM symposium on Applied computing},
  pages 2271--2278. ACM, 2008.

\bibitem[Group et~al.(2013)]{sparql11}
W3C SPARQL~Working Group et~al.
\newblock {SPARQL} 1.1 overview.
\newblock \emph{W3C Recommendation. W3C}, 2013.
\newblock http://www.w3.org/TR/sparql11-overview/.

\bibitem[Gulutzan and Pelzer(1999)]{gulutzan1999sql}
Peter Gulutzan and Trudy Pelzer.
\newblock \emph{SQL-99 complete, really}.
\newblock R \& D Books, 1999.

\bibitem[Halverson et~al.(2004)Halverson, Josifovski, Lohman, Pirahesh, and
  M{\"o}rschel]{halverson2004rox}
Alan Halverson, Vanja Josifovski, Guy Lohman, Hamid Pirahesh, and Mathias
  M{\"o}rschel.
\newblock Rox: relational over xml.
\newblock In \emph{Proceedings of the Thirtieth international conference on
  Very large data bases-Volume 30}, pages 264--275. VLDB Endowment, 2004.

\bibitem[Hausenblas et~al.(2012)Hausenblas, Grossman, Harth, and
  Cudr{\'e}-Mauroux]{hausenblas2012large}
Michael Hausenblas, Robert Grossman, Andreas Harth, and Philippe
  Cudr{\'e}-Mauroux.
\newblock Large-scale linked data processing-cloud computing to the rescue?.
\newblock \emph{CLOSER}, 20\penalty0 (2):\penalty0 246--251, 2012.

\bibitem[He and Naughton(1999)]{he1999relational}
Jayavel Shanmugasundaram Kristin Tufte~Gang He and Chun Zhang David
  DeWitt~Jeffrey Naughton.
\newblock Relational databases for querying xml documents: Limitations and
  opportunities.
\newblock \emph{Very Large Data Bases: Proceedings}, 25:\penalty0 302, 1999.

\bibitem[Hu and Chen(2008)]{hu2008adaptive}
Tian-lei Hu and Gang Chen.
\newblock Adaptive xml to relational mapping: an integrated approach.
\newblock \emph{Journal of Zhejiang University-SCIENCE A}, 9\penalty0
  (6):\penalty0 758--769, 2008.

\bibitem[Jigyasu et~al.(2006)Jigyasu, Banerjee, Borkar, Carey, Dixit, Malkani,
  and Thatte]{jigyasu2006sql}
Sunil Jigyasu, Sujeet Banerjee, Vinayak Borkar, Michael Carey, Kanad Dixit,
  Anil Malkani, and Sachin Thatte.
\newblock Sql to xquery translation in the aqualogic data services platform.
\newblock In \emph{Data Engineering, 2006. ICDE'06. Proceedings of the 22nd
  International Conference on}, pages 97--97. IEEE, 2006.

\bibitem[Jonathan~Robie(2017)]{jonathan2017xml}
Josh~Spiegel Jonathan~Robie, Michael~Dyck.
\newblock Xquery 3.1: An xml query language, 2017.
\newblock URL \url{https://www.w3.org/TR/xquery-31/}.
\newblock Accessed: 06-08-2019.

\bibitem[Kiminki et~al.(2010)Kiminki, Knuuttila, and
  Hirvisalo]{kiminki2010sparql}
Sami Kiminki, Jussi Knuuttila, and Vesa Hirvisalo.
\newblock Sparql to sql translation based on an intermediate query language.
\newblock In \emph{The 6th International Workshop on Scalable Semantic Web
  Knowledge Base Systems (SSWS2010)}, page~32, 2010.

\bibitem[Krishnamurthy et~al.(2003)Krishnamurthy, Kaushik, and
  Naughton]{krishnamurthy2003xml}
Rajasekar Krishnamurthy, Raghav Kaushik, and Jeffrey~F Naughton.
\newblock Xml-to-sql query translation literature: The state of the art and
  open problems.
\newblock In \emph{International XML Database Symposium}, pages 1--18.
  Springer, 2003.

\bibitem[Krishnamurthy et~al.(2004{\natexlab{a}})Krishnamurthy, Chakaravarthy,
  Kaushik, and Naughton]{krishnamurthy2004recursive}
Rajasekar Krishnamurthy, Venkatesan~T Chakaravarthy, Raghav Kaushik, and
  Jeffrey~F Naughton.
\newblock Recursive xml schemas, recursive xml queries, and relational storage:
  Xml-to-sql query translation.
\newblock In \emph{null}, page~42. IEEE, 2004{\natexlab{a}}.

\bibitem[Krishnamurthy et~al.(2004{\natexlab{b}})Krishnamurthy, Kaushik, and
  Naughton]{krishnamurthy2004efficient}
Rajasekar Krishnamurthy, Raghav Kaushik, and Jeffrey~F Naughton.
\newblock Efficient xml-to-sql query translation: Where to add the
  intelligence?
\newblock In \emph{Proceedings of the Thirtieth international conference on
  Very large data bases-Volume 30}, pages 144--155. VLDB Endowment,
  2004{\natexlab{b}}.

\bibitem[Laney(2012)]{laney2012deja}
Doug Laney.
\newblock Deja vvvu: others claiming gartner’s construct for big data.
\newblock \emph{Gartner Blog, Jan}, 14, 2012.

\bibitem[Lawrence(2014)]{lawrence2014integration}
Ramon Lawrence.
\newblock Integration and virtualization of relational sql and nosql systems
  including mysql and mongodb.
\newblock In \emph{Computational Science and Computational Intelligence (CSCI),
  2014 International Conference on}, volume~1, pages 285--290. IEEE, 2014.

\bibitem[Lindaaker(2018)]{lindaaker2018graphhistory}
Tobias Lindaaker.
\newblock An overview of the recent history of graph query languages, 2018.

\bibitem[Lu et~al.(2008)Lu, Cao, Ma, Yu, and Pan]{lu2008effective}
Jing Lu, Feng Cao, Li~Ma, Yong Yu, and Yue Pan.
\newblock An effective sparql support over relational databases.
\newblock In \emph{Semantic web, ontologies and databases}, pages 57--76.
  Springer, 2008.

\bibitem[Mahmoud(2017)]{topGraphDatabase}
Alaa Mahmoud.
\newblock No more joins: An overview of graph database query language, 2017.
\newblock URL
  \url{https://developer.ibm.com/dwblog/2017/overview-graph-database-query-languages/}.
\newblock Accessed: 06-08-2019.

\bibitem[Mani et~al.(2006)Mani, Wang, Dougherty, and
  Rundensteiner]{mani2006join}
Murali Mani, Song Wang, Dan Dougherty, and Elke~A Rundensteiner.
\newblock Join minimization in xml-to-sql translation: an algebraic approach.
\newblock \emph{ACM SIGMOD Record}, 35\penalty0 (1):\penalty0 20--25, 2006.

\bibitem[Melnik(2001)]{melnik2001storing}
Sergey Melnik.
\newblock Storing rdf in a relational database, 2001.

\bibitem[Michel et~al.(2014)Michel, Montagnat, and Zucker]{michel2014survey}
Franck Michel, Johan Montagnat, and Catherine~Faron Zucker.
\newblock \emph{A survey of RDB to RDF translation approaches and tools}.
\newblock PhD thesis, I3S, 2014.

\bibitem[Michel et~al.(2016)Michel, Zucker, and Montagnat]{michel2016generic}
Franck Michel, Catherine~Faron Zucker, and Johan Montagnat.
\newblock A generic mapping-based query translation from sparql to various
  target database query languages.
\newblock In \emph{12th International Conference on Web Information Systems and
  Technologies (WEBIST'16)}, 2016.

\bibitem[Min et~al.(2008)Min, Lee, and Chung]{min2008xtron}
Jun-Ki Min, Chun-Hee Lee, and Chin-Wan Chung.
\newblock Xtron: An xml data management system using relational databases.
\newblock \emph{Information and Software Technology}, 50\penalty0 (5):\penalty0
  462--479, 2008.

\bibitem[Mutharaju et~al.(2013)Mutharaju, Sakr, Sala, and
  Hitzler]{mutharaju2013d}
Raghava Mutharaju, Sherif Sakr, Alessandra Sala, and Pascal Hitzler.
\newblock D-sparq: distributed, scalable and efficient rdf query engine.
\newblock 2013.

\bibitem[Noy(2004)]{noy2004semantic}
Natalya~F Noy.
\newblock Semantic integration: a survey of ontology-based approaches.
\newblock \emph{ACM Sigmod Record}, 33\penalty0 (4):\penalty0 65--70, 2004.

\bibitem[P{\'e}rez et~al.(2006)P{\'e}rez, Arenas, and
  Gutierrez]{perez2006semantics}
Jorge P{\'e}rez, Marcelo Arenas, and Claudio Gutierrez.
\newblock Semantics and complexity of sparql.
\newblock In \emph{International semantic web conference}, pages 30--43.
  Springer, 2006.

\bibitem[P{\'e}rez et~al.(2009)P{\'e}rez, Arenas, and
  Gutierrez]{perez2009semantics}
Jorge P{\'e}rez, Marcelo Arenas, and Claudio Gutierrez.
\newblock Semantics and complexity of sparql.
\newblock \emph{ACM Transactions on Database Systems (TODS)}, 34\penalty0
  (3):\penalty0 16, 2009.

\bibitem[Poggi et~al.(2008)Poggi, Lembo, Calvanese, De~Giacomo, Lenzerini, and
  Rosati]{poggi2008linking}
Antonella Poggi, Domenico Lembo, Diego Calvanese, Giuseppe De~Giacomo, Maurizio
  Lenzerini, and Riccardo Rosati.
\newblock Linking data to ontologies.
\newblock In \emph{Journal on data semantics X}, pages 133--173. Springer,
  2008.

\bibitem[Polleres and Wallner(2013)]{polleres2013relation}
Axel Polleres and Johannes~Peter Wallner.
\newblock On the relation between sparql1. 1 and answer set programming.
\newblock \emph{Journal of Applied Non-Classical Logics}, 23\penalty0
  (1-2):\penalty0 159--212, 2013.

\bibitem[Priyatna et~al.(2014)Priyatna, Corcho, and
  Sequeda]{priyatna2014formalisation}
Freddy Priyatna, Oscar Corcho, and Juan Sequeda.
\newblock Formalisation and experiences of r2rml-based sparql to sql query
  translation using morph.
\newblock In \emph{Proceedings of the 23rd international conference on World
  wide web}, pages 479--490. ACM, 2014.

\bibitem[Prudhommeaux(2008)]{prudhommeaux2008sparql}
Eric Prudhommeaux.
\newblock Sparql query language for rdf.
\newblock \emph{http://www. w3. org/TR/rdf-sparql-query/}, 2008.

\bibitem[Prud’hommeaux(2004)]{prud2004optimal}
Eric Prud’hommeaux.
\newblock Optimal rdf access to relational databases, 2004.

\bibitem[Prud’Hommeaux et~al.(2008)Prud’Hommeaux, Seaborne,
  et~al.]{sparql10}
Eric Prud’Hommeaux, Andy Seaborne, et~al.
\newblock {SPARQL} query language for {RDF}.
\newblock \emph{W3C recommendation}, 15, 2008.
\newblock www.w3.org/TR/rdf-sparql-query/.

\bibitem[QueryMongo(2019)]{querymongo}
QueryMongo.
\newblock Query mongo: Mysql to mongodb query translator, 2019.
\newblock URL \url{http://www.querymongo.com}.
\newblock Accessed: 06-08-2019.

\bibitem[Rachapalli et~al.(2011)Rachapalli, Khadilkar, Kantarcioglu, and
  Thuraisingham]{rachapalli2011retro}
Jyothsna Rachapalli, Vaibhav Khadilkar, Murat Kantarcioglu, and Bhavani
  Thuraisingham.
\newblock Retro: a framework for semantics preserving sql-to-sparql
  translation.
\newblock \emph{The University of Texas at Dallas}, 800:\penalty0 75080--3021,
  2011.

\bibitem[Ramanujam et~al.(2009{\natexlab{a}})Ramanujam, Gupta, Khan, Seida, and
  Thuraisingham]{ramanujam2009r2d}
Sunitha Ramanujam, Anubha Gupta, Latifur Khan, Steven Seida, and Bhavani
  Thuraisingham.
\newblock R2d: A bridge between the semantic web and relational visualization
  tools.
\newblock In \emph{2009 IEEE International Conference on Semantic Computing},
  pages 303--311. IEEE, 2009{\natexlab{a}}.

\bibitem[Ramanujam et~al.(2009{\natexlab{b}})Ramanujam, Gupta, Khan, Seida, and
  Thuraisingham]{ramanujam2009r2dextra}
Sunitha Ramanujam, Anubha Gupta, Latifur Khan, Steven Seida, and Bhavani
  Thuraisingham.
\newblock R2d: Extracting relational structure from rdf stores.
\newblock In \emph{Proceedings of the 2009 IEEE/WIC/ACM International Joint
  Conference on Web Intelligence and Intelligent Agent Technology-Volume 01},
  pages 361--366. IEEE Computer Society, 2009{\natexlab{b}}.

\bibitem[Ranking(2019)]{DBEnginesRanking}
DB-Engines Ranking.
\newblock Db-engines ranking, 2019.
\newblock URL \url{https://db-engines.com/en/ranking}.
\newblock Accessed: 06-08-2019.

\bibitem[Reddy(2015)]{mongoDB-translator-teiid}
Ramesh Reddy.
\newblock Mongodb translator - teiid 9.0 (draft), 2015.
\newblock URL
  \url{https://docs.jboss.org/author/display/TEIID/MongoDB+Translator}.
\newblock Accessed: 06-08-2019.

\bibitem[Rodriguez(2015)]{rodriguez2015gremlin}
Marko~A Rodriguez.
\newblock The gremlin graph traversal machine and language (invited talk).
\newblock In \emph{Proceedings of the 15th Symposium on Database Programming
  Languages}, pages 1--10. ACM, 2015.

\bibitem[Rodriguez-Muro and Rezk(2015)]{journals/ws/Rodriguez-MuroR15}
Mariano Rodriguez-Muro and Martín Rezk.
\newblock Efficient sparql-to-sql with r2rml mappings.
\newblock \emph{J. Web Sem.}, 33:\penalty0 141--169, 2015.
\newblock URL \url{http://dblp.uni-trier.de/db/journals/ws/ws33.html#Rodriguez-
  MuroR15}.

\bibitem[Rodr{\i}guez-Muro et~al.(2012)Rodr{\i}guez-Muro, Hardi, and
  Calvanese]{rodriguez2012quest}
Mariano Rodr{\i}guez-Muro, Josef Hardi, and Diego Calvanese.
\newblock Quest: efficient sparql-to-sql for rdf and owl.
\newblock In \emph{11th International Semantic Web Conference ISWC}, page~53.
  Citeseer, 2012.

\bibitem[Russell(2016)]{sql-to-mongo-db-query-converter}
Vincent Russell.
\newblock sql-to-mongo-db-query-converter, 2016.
\newblock URL
  \url{https://github.com/vincentrussell/sql-to-mongo-db-query-converter}.
\newblock Accessed: 15-10-2018.

\bibitem[Sadalage and Fowler(2013)]{sadalage2013nosql}
Pramod~J Sadalage and Martin Fowler.
\newblock \emph{NoSQL distilled: a brief guide to the emerging world of
  polyglot persistence}.
\newblock Pearson Education, 2013.

\bibitem[Sahoo et~al.(2009)Sahoo, Halb, Hellmann, Idehen, Thibodeau~Jr, Auer,
  Sequeda, and Ezzat]{sahoo2009survey}
Satya~S Sahoo, Wolfgang Halb, Sebastian Hellmann, Kingsley Idehen, Ted
  Thibodeau~Jr, S{\"o}ren Auer, Juan Sequeda, and Ahmed Ezzat.
\newblock A survey of current approaches for mapping of relational databases to
  rdf.
\newblock \emph{W3C RDB2RDF Incubator Group Report}, 1:\penalty0 113--130,
  2009.

\bibitem[Schreiner et~al.(2015)Schreiner, Duarte, and dos
  Santos~Mello]{schreiner2015sqltokeynosql}
Geomar~A Schreiner, Denio Duarte, and Ronaldo dos Santos~Mello.
\newblock Sqltokeynosql: a layer for relational to key-based nosql database
  mapping.
\newblock In \emph{Proceedings of the 17th International Conference on
  Information Integration and Web-based Applications \& Services}, page~74.
  ACM, 2015.

\bibitem[Sequeda and Miranker(2013)]{sequeda2013ultrawrap}
Juan~F Sequeda and Daniel~P Miranker.
\newblock Ultrawrap: Sparql execution on relational data.
\newblock \emph{Web Semantics: Science, Services and Agents on the World Wide
  Web}, 22:\penalty0 19--39, 2013.

\bibitem[Spanos et~al.(2012)Spanos, Stavrou, and Mitrou]{spanos2012bringing}
Dimitrios-Emmanuel Spanos, Periklis Stavrou, and Nikolas Mitrou.
\newblock Bringing relational databases into the semantic web: A survey.
\newblock \emph{Semantic Web}, 3\penalty0 (2):\penalty0 169--209, 2012.

\bibitem[SQL2Gremlin(2018)]{sql2gremlin}
SQL2Gremlin.
\newblock Sql2gremlin, 2018.
\newblock URL \url{http://sql2gremlin.com}.
\newblock Accessed: 06-08-2019.

\bibitem[Stadler et~al.(2013)Stadler, Unbehauen, Lehmann, and
  Auer]{stadlerconnecting}
Claus Stadler, J{\"o}rg Unbehauen, Jens Lehmann, and S{\"o}ren Auer.
\newblock Connecting crowdsourced spatial information to the data web with
  sparqlify.
\newblock Technical report, Technical Report, University of Leipzig, Leipzig,
  2013. Available at
  http://sparqlify.org/downloads/documents/2013-Sparqlify-Technical-Report.pdf,
  2013.

\bibitem[Steer et~al.(2017)Steer, Alnaimi, Lotz, Cuadrado, Vaquero, and
  Varvenne]{steer2017cytosm}
Benjamin~A Steer, Alhamza Alnaimi, Marco~ABFG Lotz, Felix Cuadrado, Luis~M
  Vaquero, and Joan Varvenne.
\newblock Cytosm: Declarative property graph queries without data migration.
\newblock In \emph{Proceedings of the Fifth International Workshop on Graph
  Data-management Experiences \& Systems}, page~4. ACM, 2017.

\bibitem[Sun et~al.(2015)Sun, Fokoue, Srinivas, Kementsietsidis, Hu, and
  Xie]{sun2015sqlgraph}
Wen Sun, Achille Fokoue, Kavitha Srinivas, Anastasios Kementsietsidis, Gang Hu,
  and Guotong Xie.
\newblock Sqlgraph: an efficient relational-based property graph store.
\newblock In \emph{Proceedings of the 2015 ACM SIGMOD International Conference
  on Management of Data}, pages 1887--1901. ACM, 2015.

\bibitem[Thakkar et~al.(2017)Thakkar, Punjani, Auer, and
  Vidal]{thakkar2017towards}
Harsh Thakkar, Dharmen Punjani, S{\"o}ren Auer, and Maria-Esther Vidal.
\newblock Towards an {I}ntegrated {G}raph {A}lgebra for {G}raph {P}attern
  {M}atching with {G}remlin.
\newblock In \emph{International Conference on Database and Expert Systems
  Applications}, pages 81--91. Springer, 2017.

\bibitem[Thakkar et~al.(2018{\natexlab{a}})Thakkar, Punjani, Keswani, Lehmann,
  and Auer]{thakkar2018stitch}
Harsh Thakkar, Dharmen Punjani, Yashwant Keswani, Jens Lehmann, and S{\"o}ren
  Auer.
\newblock A {S}titch in {T}ime {S}aves {N}ine--{SPARQL} {Q}uerying of
  {P}roperty {G}raphs using {G}remlin {T}raversals.
\newblock \emph{arXiv preprint arXiv:1801.02911}, 2018{\natexlab{a}}.

\bibitem[Thakkar et~al.(2018{\natexlab{b}})Thakkar, Punjani, Lehmann, and
  Auer]{thakkar2018two}
Harsh Thakkar, Dharmen Punjani, Jens Lehmann, and S{\"o}ren Auer.
\newblock Two for {O}ne: {Q}uerying {P}roperty {G}raph {D}atabases using
  {SPARQL} via {G}remlinator.
\newblock In \emph{Proceedings of the 1st ACM SIGMOD Joint International
  Workshop on Graph Data Management Experiences \& Systems (GRADES) and Network
  Data Analytics (NDA)}, page~12. ACM, 2018{\natexlab{b}}.

\bibitem[Unbehauen and Martin(2016)]{unbehauen-semantics-2016-sparqlmap-m}
J{\"o}rg Unbehauen and Michael Martin.
\newblock Executing sparql queries over mapped document stores with
  sparqlmap-m.
\newblock In \emph{12th International Conference on Semantic Systems
  Proceedings (SEMANTiCS 2016)}, SEMANTiCS '16, Leipzig, Germany, September
  2016.

\bibitem[Unbehauen et~al.(2012)Unbehauen, Stadler, and
  Auer]{unbehauen2012accessing}
J{\"o}rg Unbehauen, Claus Stadler, and S{\"o}ren Auer.
\newblock Accessing relational data on the web with sparqlmap.
\newblock In \emph{Joint International Semantic Technology Conference}, pages
  65--80. Springer, 2012.

\bibitem[UnityJDBC(2017)]{unityjdbc}
UnityJDBC.
\newblock Unityjdbc - jdbc driver for mongodb, 2017.
\newblock URL \url{http://unityjdbc.com/mongojdbc/mongo_jdbc.php}.
\newblock Accessed: 06-08-2019.

\bibitem[Vidhya and Samuel(2009)]{vidhya2009query}
PM~Vidhya and Philip Samuel.
\newblock Query translation from sql to xpath.
\newblock In \emph{Nature \& Biologically Inspired Computing, 2009. NaBIC 2009.
  World Congress on}, pages 1749--1752. IEEE, 2009.

\bibitem[Vidhya and Samuel(2010)]{vidhya2010insert}
PM~Vidhya and Philip Samuel.
\newblock Insert queries in xml database.
\newblock In \emph{Computer Science and Information Technology (ICCSIT), 2010
  3rd IEEE International Conference on}, volume~1, pages 9--13. IEEE, 2010.

\bibitem[Website(2019)]{neo4jsql}
Neo4j Website.
\newblock For relational database developers: A sql to cypher guide, 2019.
\newblock URL \url{https://neo4j.com/developer/guide-sql-to-cypher}.
\newblock Accessed: 06-08-2019.

\bibitem[Wilmes(2016)]{sqlgremlin}
Ted Wilmes.
\newblock Sql-gremlin, 2016.
\newblock URL \url{https://github.com/twilmes/sql-gremlin}.
\newblock Accessed: 06-08-2019.

\end{thebibliography}
\bibliographystyle{plainnat}

\end{document}